\documentclass[iop]{emulateapj}
\usepackage{verbatim}
\usepackage{hyperref} % for \usepackage{breakurl}
\usepackage[hyphenbreaks]{breakurl} % for \burl
\usepackage[normalem]{ulem} % for strikethrough
\usepackage{cancel} % to strikethrough equations
\newcommand{\rearth}{R$_{\earth}$}
\newcommand{\rp}{$R_p$}
\newcommand{\rpasq}{$(R_p/a)^2$}

\shortauthors{Sheets \& Deming}
\shorttitle{Average Albedos of Small, Hot LC {\it Kepler} Candidates}
\begin{document}
\title{Average Albedos of Close-in Super-Earths and Neptunes from Statistical Analysis of Long Cadence {\it Kepler} Secondary Eclipse Data}

\author{Holly A. Sheets\altaffilmark{1,2,3} and Drake Deming\altaffilmark{2,3}}
\altaffiltext{1}{now at Department of Physics, McGill University, 3600 rue University, Montr\'eal, QC, H3A 2T8, CAN}
\altaffiltext{2}{Department of Astronomy, University of Maryland, College Park, MD 20742-2421}
\altaffiltext{3}{NASA Astrobiology Institute's Virtual Planetary Laboratory}
\email{holly.sheets@mcgill.ca}


\begin{abstract}
We present the results of our work to determine the average albedo for small, close-in planets in the {\it Kepler} candidate catalog.  We have adapted our method of averaging short cadence light curves of multiple {\it Kepler} planet candidates to long cadence data, in order to detect an average albedo for the group of candidates.  Long cadence data exist for many more candidates than the short cadence, and so we separate the candidates into smaller radius bins than in our previous work:  1-2 \rearth, 2-4 \rearth, and 4-6 \rearth.  We find that on average, all three groups appear darker than suggested by the short cadence result, but not as dark as many hot Jupiters.  The average geometric albedos for the three groups are 0.11 $\pm$ 0.06, 0.05 $\pm$ 0.04, and 0.11 $\pm$ 0.08, respectively, for the case where heat is uniformly distributed about the planet.  If heat redistribution is inefficient, the albedos are even lower, since there will be a greater thermal contribution to the total light from the planet.  We confirm that newly-identified false positive Kepler Object of Interest (KOI) 1662.01 is indeed an eclipsing binary at twice the period listed in the planet candidate catalog.  We also newly identify planet candidate KOI 4351.01 as an eclipsing binary, and we report a secondary eclipse measurement for Kepler-4b (KOI 7.01) of $\sim$ 7.50 ppm at a phase of $\sim$ 0.7, indicating that the planet is on an eccentric orbit.  
\end{abstract}

\section{Introduction}
Secondary eclipses of a transiting planet occur when the planet passes behind the star, resulting in a slight decrease in the total light from the system.  Secondary eclipses are often measured in the infrared, where the planet-to-star contrast is optimal due to infrared emission from the planet \citep[e.g.][]{ira,irb}.  In the {\it Kepler} bandpass, however, reflected light dominates.  The reflected light signal is proportional to the square of the ratio of the planet radius to its semi-major axis \rpasq, which is, at best, tens of parts per million (ppm), but more typically a few ppm or smaller.   This is difficult to detect for most individual candidates, even in the {\it Kepler} data \citep[e.g.][]{kep10, coughlin, esteves, angerhausen, demory,esteves2015}.  Detecting reflected light provides insight into the atmosphere and/or surface by constraining the albedo.  In \citet{paperi}, henceforth Paper I, we demonstrated that short cadence {\it Kepler} light curves of many individual candidates can be co-added constructively to detect an average secondary eclipse for the group of planets.  

In this paper, we adapt the statistical method of Paper I to the {\it Kepler} long cadence data.  The two main challenges presented by the long cadence data are the distortion of the eclipse by convolution with the 30-minute integration time for each observed data point, and also the subsequent reduction in the number of data points per eclipse compared to the short cadence data.  The long cadence data, however, offer a larger sample of planet candidates, allowing us to group the planets into more restrictive and more physically meaningful radius bins than the broad 1 to 6 R$_{\earth}$ bin of Paper I.

In Section 2, we describe the {\it Kepler} candidate data selection.  In Section 3, we discuss the modifications to the data processing and averaging method for the long cadence data.  In Section 4, we describe how we modify the thermal plus reflected light models for the long cadence data.  Section 5 considers the potential effects of false positives on our results, and Section 6 presents the results for super-Earths (1-2 \rearth), mini-Neptunes (2-4 \rearth), super-Neptunes (4-6 \rearth), as well as a discussion which explores the implications of the albedo results.  Lastly, Section 7 provides a summary.

\section{{\it Kepler} Data and Candidate Selection} % section 2
The number of planets confirmed by follow-up observation (985) is a small fraction of the total planet candidates identified (4696) by the {\it Kepler} mission \citep{dr24}.   To provide a larger sample, we therefore select our targets from the planet candidates.  The false positive rate for objects labeled as planet candidates is low, near 10 percent \citep[e.g.][]{desertfpr,fressin}, so we expect that the benefit of additional targets to increase our signal to noise will outweigh the potential interference of false positives.  In Section 5, we discuss scenarios in which false positives may influence our results.

\begin{figure}
\epsscale{1.0}
\plotone{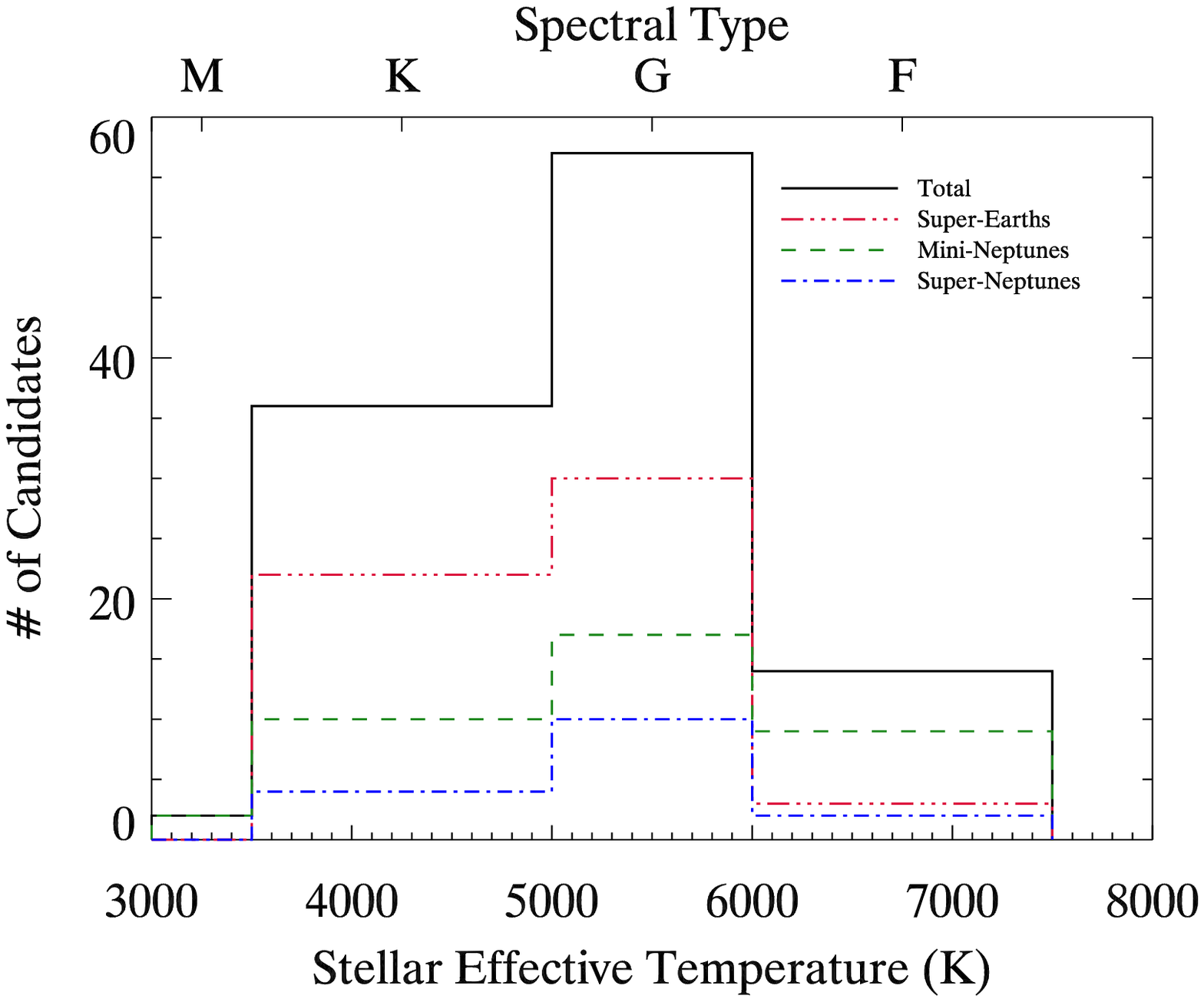}
%\vspace{1.0 in}
\caption{The spectral types of the host stars for each radius group.  Super-Earths are in red, mini-Neptunes in green, and super-Neptunes in blue.  Also shown is the total of the 3 groups in black.  Most of the host stars are G and K stars.  The two M stars in the sample host mini-Neptunes.}
\label{fig:teff_hist}
\end{figure}

\begin{figure}
\epsscale{1.0}
\plotone{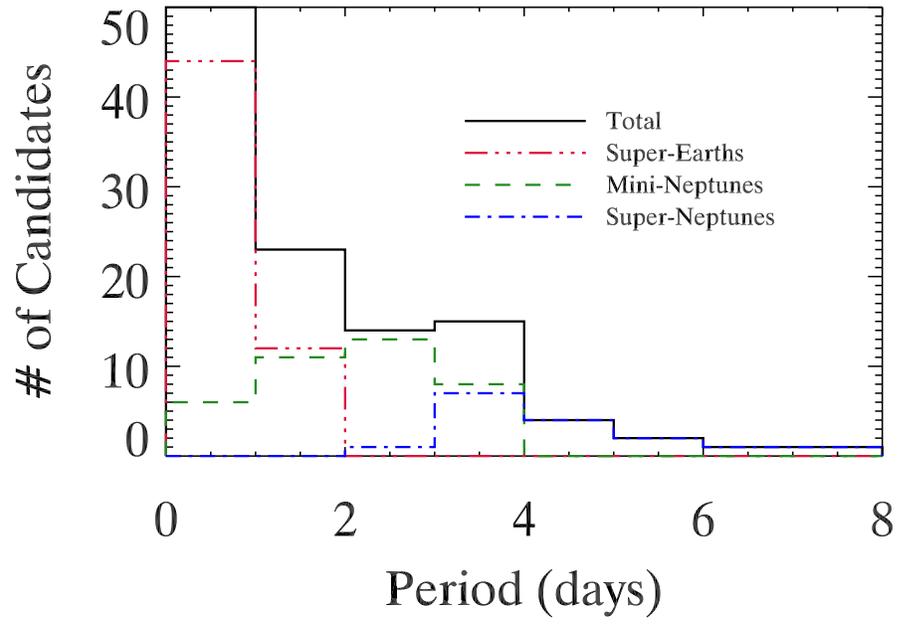}
\caption{The orbital periods of the selected candidates.  Super-Earths are shown in red, mini-Neptunes in green, and super-Neptunes in blue.  The sum of the 3 groups is shown in black.}
\label{fig:period}
\end{figure}

We select candidates with radii of 1 to 2 \rearth, 2 to 4 \rearth, and 4 to 6 \rearth~from the cumulative table of the NASA Exoplanet Archive's Kepler Objects of Interest catalog\footnote{http://exoplanetarchive.ipac.caltech.edu/}, downloaded on 2015 February 23.  We further limit the selection to candidates in these radius bins with $(R_p/a)^2 \geq 10$ ppm, where \rp~is the planet radius and $a$ is the semi-major axis of the orbit.  The expected depth of the secondary eclipse, assuming reflected light only, is the geometric albedo times this value, so we are selecting candidates that are most likely to be detectable in aggregate.  We exclude candidates with grazing transits (i.e. impact parameter $b \geq 1-(R_p/R_*)$, where \rp~is the planet candidate radius and $R_*$ is the host star radius).   We adopt circular orbits, with the center of the transit at phase 0.0 and the center of the secondary eclipse at phase 0.5.  We address the possibility of non-circular orbits in the discussion in Section 5.4.  The KOI catalog has evolved while this manuscript was in preparation.  Using the 2015 February download of the KOI cumulative table, our selection criteria combined with the tests performed in Section 3.1 result in 56 super-Earth (1-2 \rearth) candidates, 38 mini-Neptune (2-4 \rearth) candidates, and 16 super-Neptune (4-6 \rearth) candidates, whose parameters are given in Table \ref{tab:param1} and Table \ref{tab:param2}.   The various catalogs that feed into the cumulative table were finalized late in 2015, resulting in some of these objects changing status to false positives.  Additionally, \citet{fpp} provided false positive probabilities for nearly all of the candidates in the cumulative table.  To demonstrate the robustness of our results against false positives, we present in Section 6 both the results using the 2015 February catalog, as well as the results using subsets of the 3 groups, from which we have removed planet candidates with a false positive probability greater than 1\% and a few newly-identified false positives.  The shortened groups contain 40 super-Earth candidates, 28 mini-Neptune candidates, and 12 super-Neptune candidates.  We have also removed Kepler-4b from the shortened super-Neptune group, because it appears to be on an eccentric orbit (see Section 3.2.1).  The objects removed are identified in Table \ref{tab:param1} and Table \ref{tab:param2} with a footnote.  Most of the host stars in our sample are sun-like or nearly so;  The distribution of spectral types and stellar effective temperatures is given in Figure \ref{fig:teff_hist} for the 2015 February version of the groups.  Because of our $(R_p/a)^2 \geq 10$ ppm selection criterion, our sample is composed of very short-period planets.  The distribution of orbital periods for the 2015 February groups is shown in Figure \ref{fig:period}, covering a range from about 0.49 to 7.3 days.

We use the long-cadence ($\approx$ 30 min exposure) Pre-search Data Conditioning (PDC) data from the Mikulski Archive for Space Telescopes (MAST\footnote{http://archive.stsci.edu/kepler/}), for quarters 0 through 17.  We first eliminate statistical outliers from the photometric data by removing any point that is more than three times the photometric error away from the median value of the point and the four points to each side.  We normalize the light curves by dividing the photometric time series for each quarter by its mean value, and then subtracting 1 and multiplying by 10$^6$ to convert to ppm.  We mask any transits of other known objects in the system.  We select data within $\pm$ 12 hours of each eclipse time, with two restrictions, and we fit a quadratic baseline to the portion of those data that lie outside of the eclipse.  For candidates with transits (and, therefore, eclipses) whose duration is greater than 4 hours, we widen the data selection window from 12 hours to 3 times the duration of the candidate's transit.  This provides a better baseline for longer-duration eclipses.  For candidates with orbital periods around 1 day or less, the 12 hour selection window around the eclipse center would include the transits, so we also adapt the data selection window for these candidates to be short enough to avoid the transits.  We then subtract the quadratic fit, and repeat the process for all eclipses of the candidate in each photometric data file.  For the long cadence data, each file spans one observing quarter, which is about 3 months.  We eliminate any eclipses that do not have at least 1 point whose midpoint of the exposure time falls between second and third contact, and we also eliminate eclipses that do not have at least 8 data points before eclipse and at least 8 data points after eclipse.   We eliminate additional statistical outliers from the photometric data by removing any point that is more than three times the standard deviation of the data points around the quadratic fit.

\section{Averaging the Long Cadence Data} % section 3
\subsection{Scaling and Stacking the Individual Eclipses}
\begin{figure}
\epsscale{1.0}
\plotone{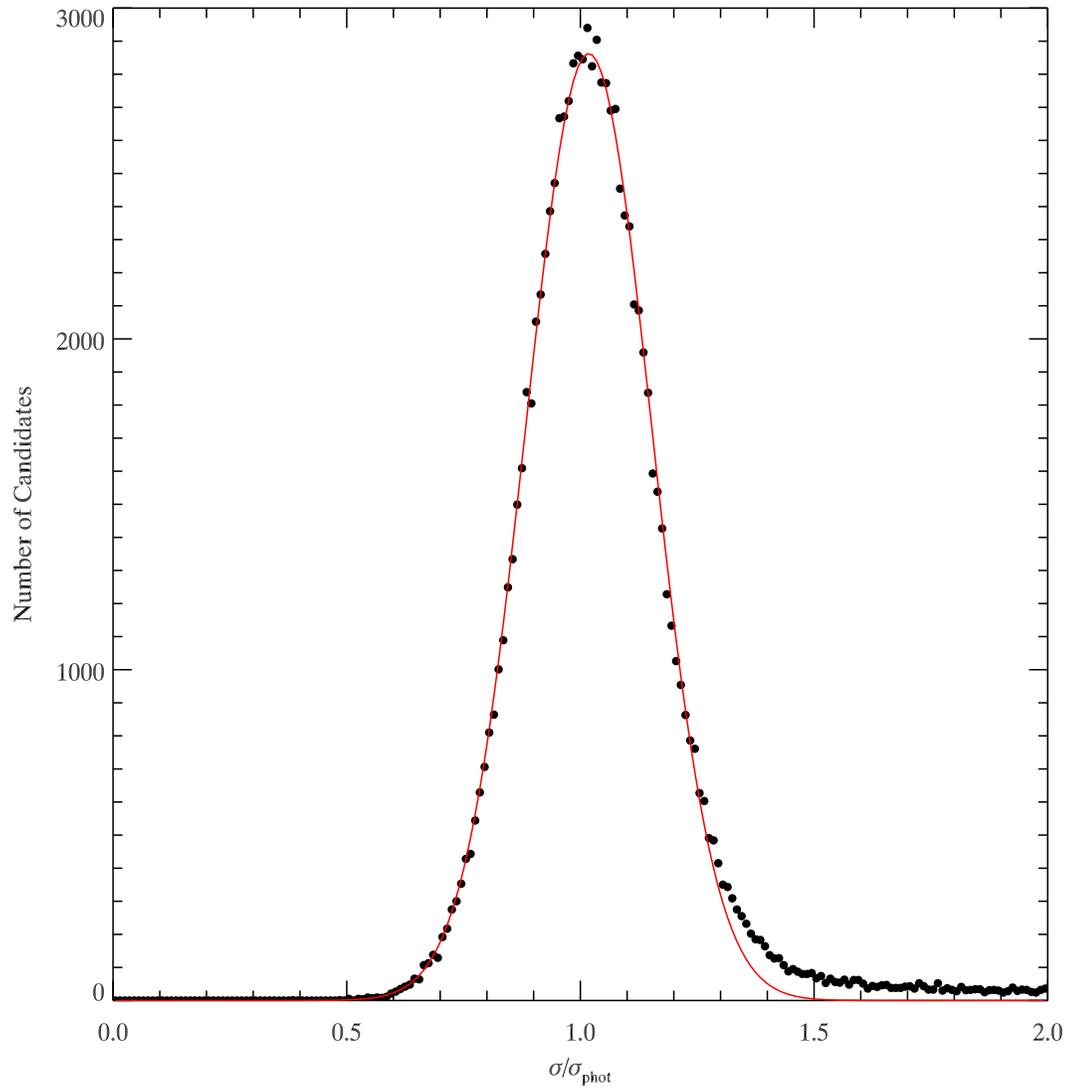}
%\vspace{1.0 in}
\caption{Shown is the histogram of the ratio of the scatter about the quadratic fit ($\sigma$) to the photometric noise ($\sigma_{phot}$) for all eclipses.  We use a cut-off of 1.3 for this parameter, keeping all eclipses with a value less than the cut-off.}
\label{fig:basesig}
\end{figure}

\begin{figure}
\epsscale{1.0}
\plotone{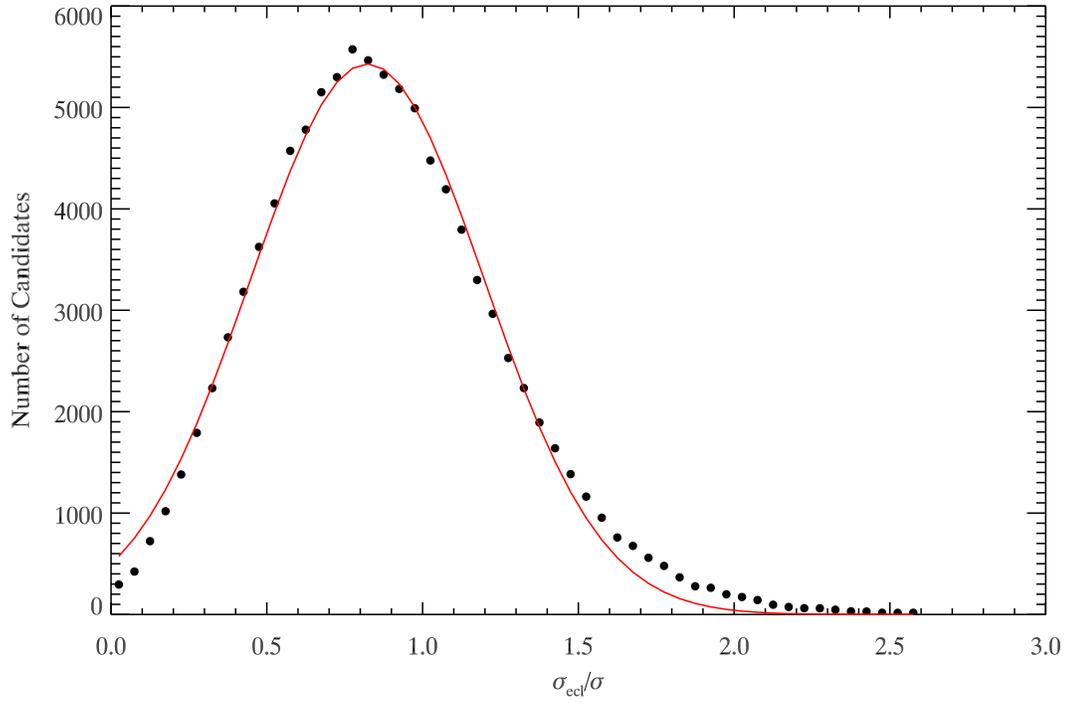}
%\vspace{1.0 in}
\caption{Shown is a histogram of the ratio of the scatter about the quadratic fit for the in-eclipse points only ($\sigma_{ecl}$) over the scatter about the quadratic fit for all of the points ($\sigma$).  We use a cut-off of 2.0 for the parameter for the groups using the 2015 February catalog, keeping all eclipses with a value less than the cut-off.   We use a slightly more restrictive cut-off of 1.5 for the results using the shortened groups.}
\label{fig:eclsig}
\end{figure}

We apply four tests to individual eclipses for each object in each group to eliminate excessive noise or instrument artifacts.  The 30 min integration time of the long cadence data means that there are fewer data points per eclipse, thus the red noise test used for the short cadence data in Paper I cannot be applied.  In its place, we use two tests.  First, we check that the standard deviation of the points ($\sigma$) around the quadratic fit from Section 2 is consistent with the mean photometric noise ($\sigma_{phot}$).  The mean photometric noise is the average of the propagated photometric errors for the data points.  We record the value of $\sigma/\sigma_{phot}$ for every eclipse, whether it passes or fails, for all of our selected candidates, and plot a histogram of the values.  The histograms of these values are consistent with a Gaussian distribution until a ratio of about 1.3, shown in Figure \ref{fig:basesig}, so we eliminate any eclipse with a ratio greater than 1.3.  We also check that the standard deviation of the points around the quadratic fit inside eclipse ($\sigma_{ecl}$) is consistent with the overall standard deviation of the points ($\sigma$).  The histograms of the ratio $\sigma_{ecl}/\sigma$ begin to deviate from Gaussian at a value of about 1.5, shown in Figure \ref{fig:eclsig}.  For the 2015 February versions of the 3 radius groups, we allowed values up to 2, but for the shortened groups we used a cut-off of 1.5.  Note that the typical photometric errors for a single eclipse are much larger than the eclipse signatures for these candidates, so this test should not be eliminating the signal of interest (see Sections 3.2.5-6).

\begin{figure}
\epsscale{1.0}
\plotone{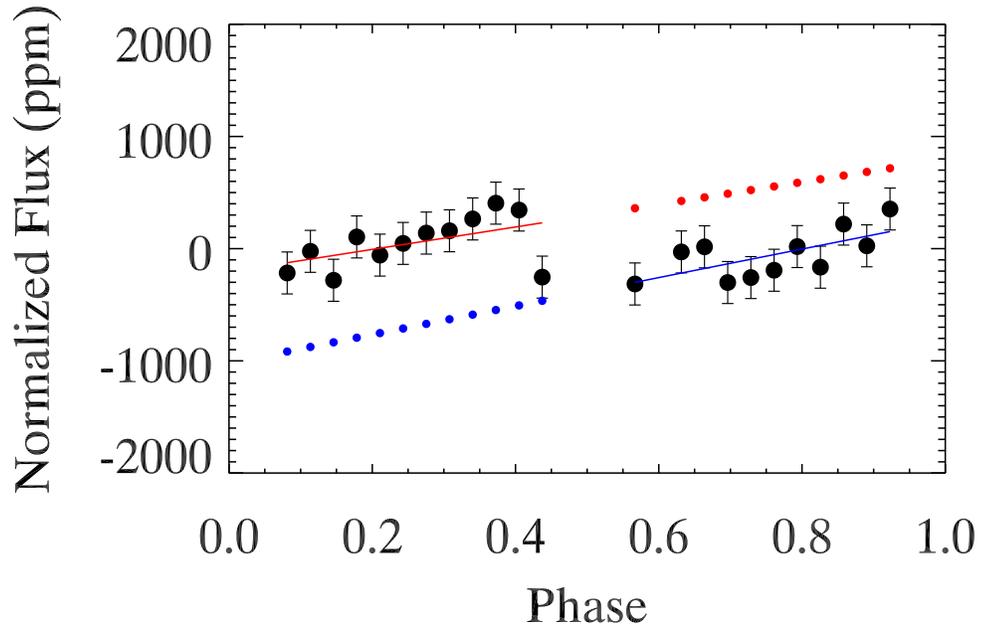}
\caption{Shown is a single eclipse from KOI 1300.01 which fails the long cadence projection test.  In the test, a line (in red) is fit to the data points (in black, with error bars) on the left (before ingress) and extrapolated (red points) for the data points (again, in black with error bars) on the right.  Similarly, a line (in blue) is fit to the black points on the right (after egress) and extrapolated (blue points) for the black points on the left.   For each side of the eclipse, the mean of the extrapolated points is calculated, as is the mean of the data points.  If the mean of the extrapolated points differs from the mean of the data points, on one side or the other, by more than three times the mean photometric errors, the eclipse is excluded.}
\label{fig:1300_bad}
\end{figure}

The third test we perform is a modification of the projection test from Paper I.  This test is applied to the data after the subtraction of the quadratic fit from Section 2.  An example of an eclipse which fails this test is given in Figure \ref{fig:1300_bad}.  We take a linear fit to the data points before ingress and use that fit to calculate a value for each of the data points after egress.  Previously we considered only the difference between the mean value of the calculated points after egress and the mean value of the actual data points after egress.  Now we instead eliminate eclipses where the difference is more than three times the mean photometric noise of the data (i.e. more than 3$\sigma$).  We also fit a line to the points after egress and project that line to the points before ingress.  The eclipse fails the test if either of the two projections fail to match the data.

The fourth test we perform is the slope test from Paper I.  We use the slopes from the linear fits in the projection test and check that they are both consistent with zero.  Any phase curve variation would be well below the photometric noise of an individual eclipse, so no slope should be detectable.  Individual data points for the brightest star in our sample, Kepler-10, have an average photometric uncertainty of 37.7 ppm, and the measured phase curve for Kepler-10b is between 7 and 10 ppm, over the range of a full orbit.  We are looking at only a fraction of the orbit, so the signal would be even smaller.  The largest phase curve amplitude expected for any planet in our sample, assuming reflected light with a (likely unrealistic) geometric albedo of 1, is 64.88 ppm (KOI-1988.01).  If the slope of both of the fits before ingress and after egress are within 3$\sigma$ of zero, the eclipse is retained.

We track the number of eclipses kept for each candidate as well as the number of eclipses rejected by each of the four tests above.  Candidates with more total rejected eclipses than retained eclipses are eliminated from the groups entirely.  We eliminate four super-Earths (KOIs 1957.01, 2324.01, 2548.01, and 3204.01), three mini-Neptunes (KOIs 2035.01, 2276.01, and 2678.01), and four super-Neptunes (KOIs 3.01, 1779.01, 1803.01, and 1804.01).

To average the light curves constructively, we use the same equations as Section 3 of Paper I to transform the phase before binning the data.  Transforming the phase ensures that we are adding data points inside of eclipse for one object to points inside of eclipse for another object, and likewise for points outside of eclipse.  Though we use points at phase outside of the 0.25 to 0.75 range to normalize the data, we only average points inside of this range.  For our assumption of circular orbits, phases of 0.25 and 0.75 are the inflection points of the phase curve, so we select these points as limits.  The only modifications we make to accommodate the long cadence data are to remove the separate ingress and egress bins that we utilized for the short cadence data and to reduce the number of bins within eclipse to 7 from 11.  Ingress and egress are typically quite short compared to the 30 min cadence, so separate bins for these two stages are no longer useful.

As we stack the eclipses of all of the candidates in each group, we also check the stacked eclipses for each candidate alone for significant secondary eclipses.  We eliminate candidates with a significant secondary eclipse (i.e. greater than 3-$\sigma$ detection) if the depth of the eclipse is incompatible with the planet interpretation, such as for KOIs 4294.01 and 4351.01 (see Sections 3.2.5-6).

\subsection{Notes on Individual Candidates}
We first constructed lists for the three radius bins from the cumulative table downloaded from the NASA Exoplanet Archive on 2014 August 27.  The various versions of the catalog that are combined to create the cumulative table were not all finalized at that time.  Many of the candidates discussed in this subsection came to our attention either while working with the 2014 August version of the cumulative table, or while comparing the 2014 August version, the downloaded version from 2015 February 23, and the now-finalized versions of the catalogs through the Q1-17 DR 24 release \citep{dr24}.   We comment here on the objects that we found notable, as guidance for other investigations.

\subsubsection{KOI 7.01 (Kepler-4b)}
\begin{figure}
\epsscale{1.0}
\plotone{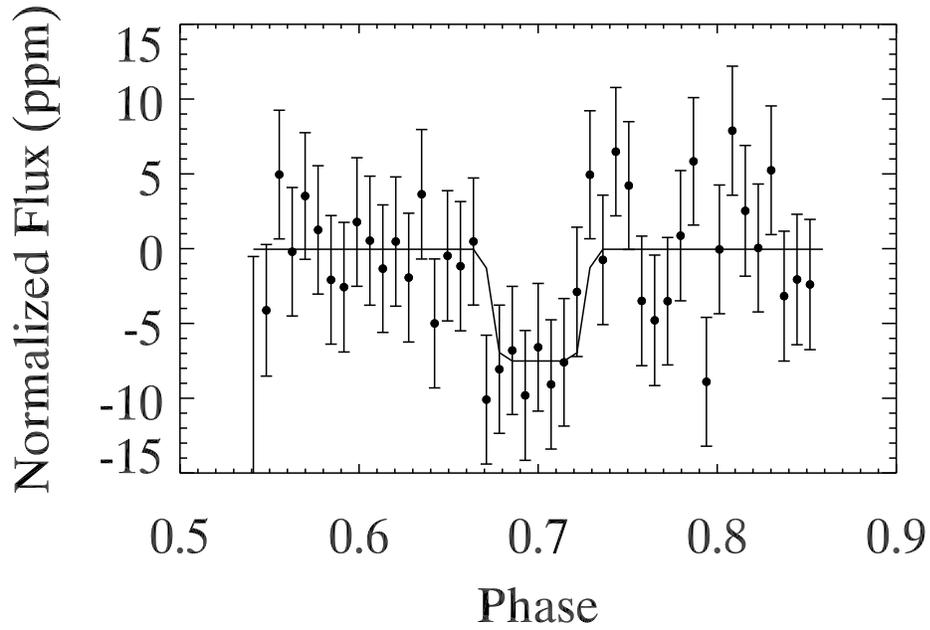}
\caption{Shown is the phase-folded light curve for KOI-7.01 (Kepler-4b) around phase 0.70, where we have detected a possible secondary eclipse.  The binned data are plotted as points, with the error bars set by propagating the photometric uncertainties.  The model, plotted as a solid black line, naively assumes an eclipse duration equal to that of the transit.  The eclipse depth from this fit is 7.47 $\pm$ 1.82 ppm, with the out-of-eclipse level at -0.039 $\pm$ 0.724 ppm.}
\label{fig:kep4b}
\end{figure}

Radial velocity data for Kepler-4b hinted that the orbit may be eccentric.  \citet{bor4b}, in the discovery paper, found that an eccentricity of 0.22 $\pm$ 0.08 fit the RV data, but at such a slight improvement of reduced $\chi^2$ that the authors adopted the circular solution.  \citet{kip4b} also found ambiguous evidence for a eccentricity of 0.208 $\pm$ 0.104 or 0.25$^{+ 0.11}_{- 0.12}$, depending on the model used for the fit.  In both studies, as well as a subsequent search by \citet{angerhausen}, the secondary eclipse was modeled only for the circular solution and not detected.  We find evidence for a secondary eclipse, with depth of 7.47 $\pm$ 1.82 ppm, centered at roughly phase 0.7 and using linear regression to fit the model to the data.  The binned data and the model are shown in Figure \ref{fig:kep4b}.  The model naively assumes the eclipse has the same duration as the transit.  Very naively, we can estimate the range of geometric albedo required for this eclipse depth, assuming the 95\% confidence upper limit of $e <$ 0.43 from \citet{kip4b}, the \rpasq= 16.13 ppm calculated from the Exoplanet Archive table, and the periastron and apastron distances, $r = a(1-e^2)/(1\pm e)$, from \citet{murray}.  These roughly translate to a geometric albedo, assuming only reflected light, of 0.15 if the eclipse occurs at apastron, and 0.95 if the eclipse occurs at periastron.  More detailed modeling of the system is in progress, but it is beyond the scope of the current work.  Kepler-4b is included in the 2015 February group, but we have removed it from the shortened group of super-Neptunes.

\subsubsection{KOI 1662.01}
\begin{figure}
\epsscale{1.0}
\plotone{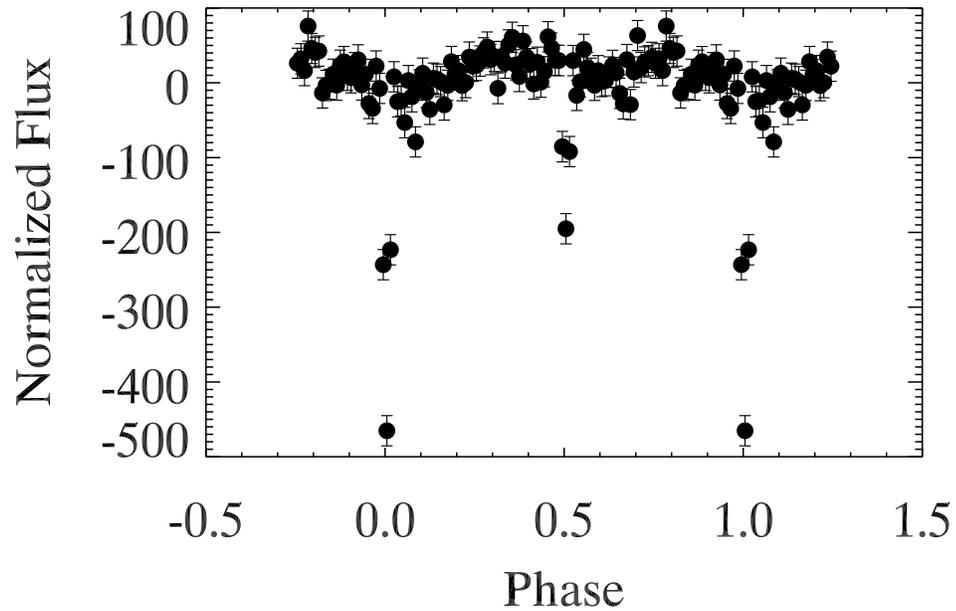}
%\vspace{1.0 in}
\caption{The light curve for KOI 1662.01, phase-folded at twice the period listed in the current KOI catalog.  This KOI is now listed as a false positive for a significant secondary eclipse.  We confirm that the secondary eclipse is significant as well as different in depth than the primary eclipse, making it clearly an eclipsing binary.}
\label{fig:1662}
\end{figure}

KOI 1662.01 was listed as a 1.66 \rearth~candidate in the Q1-12 catalog with a period of 0.784 days.  It is now listed as a false positive in the Q1-17 catalog due to a significant secondary eclipse.  The period is still listed as 0.784 days, but the true period is twice this value.  The Q1-17 DR25 TCE (Threshold-Crossing Events) table at the NASA Exoplanet Archive lists two ``planets'' for this star:  one at the 0.784 day period and one at twice that period.  Figure \ref{fig:1662} shows the data phase-folded at the 1.568 day period, with a significant secondary eclipse, indicating that it is a blended eclipsing binary.  It is included as a super-Earth in the tests discussed in Section 5 as well as the 2015 February group in Section 6, but removed from the shortened group of super-Earths.   

\subsubsection{KOIs 2396.01 and 2882.02}
KOIs 2396.01 and 2882.02 are also now listed in the Q1-17 catalog as false positives for significant secondaries.  We retain these two candidates in our shortened group of super-Earths because we do not see a secondary eclipse when folding at their planet candidate periods, nor do we see a significant difference between the ``primary'' and ``secondary'' eclipse depths when folding at twice the listed period.  Furthermore, they have less than a 1\% false positive probability in \citet{fpp}, unlike KOI 1662.01.

\subsubsection{KOI 2795.01}
\begin{figure}
\epsscale{1.0}
\plotone{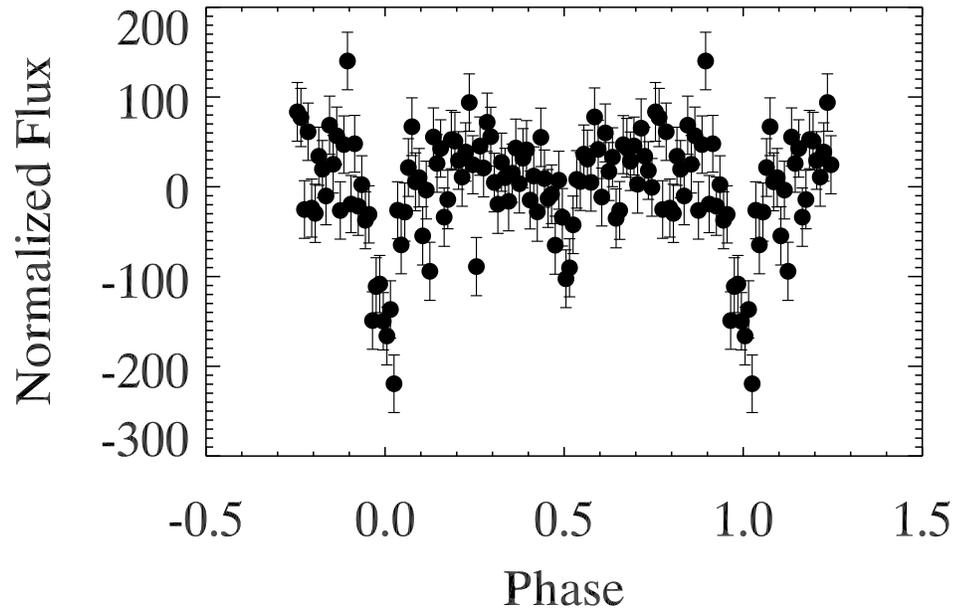}
%\vspace{1.0 in}
\caption{Averaged light curve for KOI 2795.01 from Q4, Q5, Q8, Q9, Q12, Q13, Q16, and Q17, where only a weak signal is present.  Here we see a significant secondary eclipse at phase 0.5.}
\label{fig:2795_weak}
\end{figure}

\begin{figure}
\epsscale{1.0}
\plotone{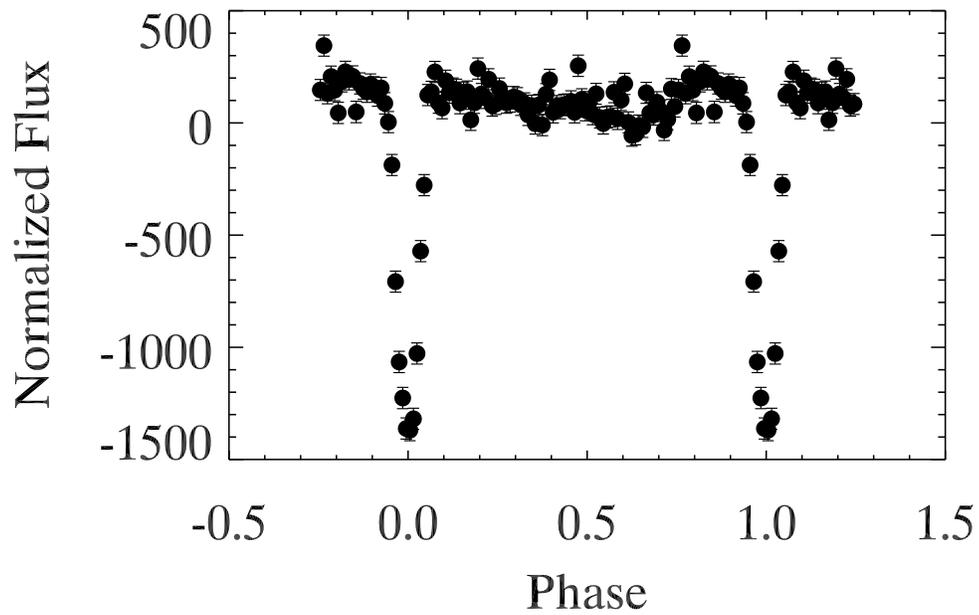}
%\vspace{1.0 in}
\caption{Averaged light curve for KOI 2795.01 from Q6, Q10, and Q14, where a strong signal is present.  Here we see a much stronger primary eclipse at phase 0.0 (and repeated at phase 1.0) than the primary eclipse in Figure \ref{fig:2795_weak}.}
\label{fig:2795_strong}
\end{figure}

KOI 2795.01 is listed as a 2.4 \rearth~candidate in the Q1-8 catalog and as a false positive due to a significant secondary eclipse in the Q1-17 DR 24 catalog.  We find that the signal strength for KOI 2795.01 changes depending on the orientation of the telescope.  It appears strongly in Q6, Q10, and Q14, and to a lesser extent in Q7, Q11, and Q15.  There are no data for Q0-3, and the signal appears only weakly in the remaining quarters.  Furthermore, the shape of the light curve changes when comparing the weak signal quarters, shown in Figure \ref{fig:2795_weak}, to the strong signal quarters, shown in Figure \ref{fig:2795_strong}.  The spacecraft rolls by 90 degrees at the end of each quarter, and so on every fourth quarter, a target returns to the same CCD.  This correlation with certain orientations of the telescope suggests contamination by electrical cross-talk, a column anomaly, or internal reflections \citep{contam}.  It is included in our tests discussed in Section 5 and our 2015 February mini-Neptunes group in Section 6, but it is not included in the shortened group of mini-Neptunes.

\subsubsection{KOI 4294.01}
KOI 4294.01 was listed as a 5.02 \rearth~candidate in the cumulative KOI table downloaded on 2014 August 27.   We detected a significant secondary eclipse that was much too deep to be the secondary eclipse of a planet.  It was independently found to have a significant secondary in the Q1-17 DR 24 catalog and is now listed as a false positive.   We do not include it in either of our super-Neptune groups.

\subsubsection{KOI 4351.01}
\begin{figure}
\epsscale{1.0}
\plotone{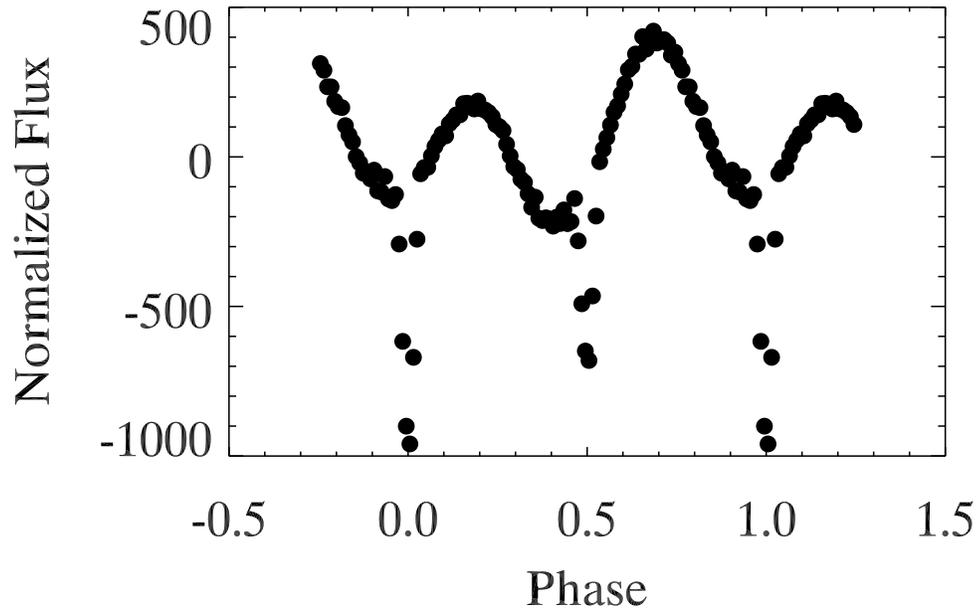}
%\vspace{1.0 in}
\caption{Averaged light curve for KOI 4351.01, which is listed in the KOI catalog as a planet candidate.  It shows a significant secondary eclipse at phase 0.5 and so must be an eclipsing binary.}
\label{fig:4351}
\end{figure}

KOI 4351.01 was listed as a 5.47 \rearth~candidate in the cumulative KOI table downloaded on 2014 August 27.    Like KOI 4294.01 above, we detected a significant, non-planetary secondary eclipse while screening the candidates to be used in our average, indicating the object is actually an eclipsing binary.  Unlike KOI 4294.01, this candidate was not re-evaluated in the Q1-17 DR 24 catalog (Coughlin, private communication).  It is listed in the finalized Q1-12 catalog and the current cumulative table as a 24.4 \rearth~candidate.  The secondary eclipse is shown in Figure \ref{fig:4351}.   We do not include it in either of our super-Neptune groups.

\subsubsection{KOI 4924.01}
KOI 4924.01 was listed as a 3.65 \rearth~candidate in the cumulative KOI table downloaded on 2014 August 27.  Subsequently, it was listed as a false positive in the cumulative KOI table downloaded on 2015 February 23.  In the finalized Q1-16 catalog, it is still listed as a false positive due to a centroid offset flag, but in the finalized Q1-17 DR 24 catalog, it is listed as a 3.49 \rearth~candidate.   We do not include it in either of our mini-Neptune groups.

\section{Modeling Reflected Light and Thermal Emission} % section 4
As in Paper I, the reflected light is given by:
\begin{equation}\label{eq:refldepth}
\frac{F_p}{F_*}=A_g\left(\frac{R_p}{a}\right)^2
\end{equation}
where $F_p$ is the flux of reflected light from the planet, $F_*$ is the flux from the star, $A_g$ is the geometric albedo, $R_p$ is the planet radius, and $a$ is the orbital distance of the planet.  

To model the thermal emission, as in Paper I, we estimate the planet's effective temperature using the relation:
\begin{equation}
T_p = T_*\left(\frac{R_*}{a}\right)^{1/2}[f(1-A_B)]^{1/4}
\end{equation}
where $T_p$ is the planet's temperature, $T_*$ is the star's effective temperature, $a$ is the planet's orbital distance, $R_*$ is the stellar radius, and $A_B$ is the Bond albedo.  The stellar parameters are taken from the update to the Kepler Input Catalog (KIC) by \citet{2014update}.  The redistribution factor $f$ typically varies between a value of 1/4 for complete redistribution of heat around the planet to 2/3 for instantaneous re-radiation of heat \citep[see, e.g.][]{hansen,esteves,lopez,rowe06}.  We also adopt $A_B = (3/2)A_g$, which is the relation for a Lambertian surface.  We assume the thermal emission from the planet is a blackbody with this effective temperature, integrated over the bandpass of {\it Kepler} using its transmission function\footnote{http://keplergo.arc.nasa.gov/kepler\_response\_hires1.txt}.  We calculate the planet's intensity for the two extremes of $f$.  We integrate an ATLAS\footnote{http://kurucz.harvard.edu/} model atmosphere \citep{kurucz} over the bandpass for the appropriate stellar parameters to calculate the stellar intensities.  We divide the planet's intensity by the stellar intensity and scale by (\rp/$R_*$)$^2$.   

The combination of the thermal emission and reflected light gives the expected depth of the secondary eclipse.  We multiply this depth by a \citet{agol} model, with limb darkening set to zero, for the eclipse, which has been rescaled so that the continuum is 0 and full eclipse is -1.  To account for the 30 min cadence of the long cadence data, we oversample the \citet{agol} model by 0.0001 times the period of the candidate and average it over the 30 min exposure corresponding to each data point before multiplying by the expected depth for the thermal emission plus reflected light.  The periods in our candidate sample range from about 0.49 days to 7.26 days, which translates to a sampling rate of the model of 4.26 seconds to 62.7 seconds.  We then build up a thermal emission and reflected light model for the averaged data by generating a model point for each data point and averaging the model points just as the data are averaged.  To determine the average geometric albedo, we iteratively adjust the geometric albedo used to calculate the model until the averaged model matches the fitted depth of the averaged data.  %0.4934 days to 7.2584 days

\section{False Positive and Other Detrimental Scenarios} % section 5
In anticipation of our results, we here discuss possible scenarios that could falsely drive down our measured albedos.

\subsection{Dilution by Unknown Objects in Aperture}
We considered the possibility that some of the host stars in our candidate sample may have a close, unresolved companion star in the {\it Kepler} data, due to the large pixel scale of 3.98 arcsec pixel$^{-1}$ \citep{pixel}.  Unresolved companions dilute the signal from the planet in transit and in eclipse, making the planet seem smaller than it truly is.  This possible dilution means that some of our candidates may be in the wrong radius bin.  Even if the inferred radius of a candidate is wrong, however, the geometric albedo calculated from the candidate is unaffected.  

Dilution diminishes the depth of the transit as well as the depth of the secondary eclipse, but the ratio of these two depths remains the same.  The depth of the transit is given by:
\begin{equation}\label{eq:transitdepth}
\frac{F_t}{F_*} = \left(\frac{R_p}{R_*}\right)^2
\end{equation}
where $F_t$ is the flux blocked by the planet during transit.  The secondary eclipse depth, in the simplifying assumption that the eclipse is due only to reflected light, is given in Equation \ref{eq:refldepth}.  Taking the ratio ($r_{e/t}$) of the secondary eclipse depth to the transit depth gives:
\begin{equation}\label{eq:ratio}
r_{e/t} = A_g\left(\frac{R_*}{a}\right)^2
\end{equation}
Solving Equation \ref{eq:ratio} for the geometric albedo gives:
\begin{equation}
A_g = r_{e/t}\left(\frac{a}{R_*}\right)^2
\end{equation}
which does not depend on the inferred radius of the planet from the transit depth.

Although dilution by companion stars does not directly bias our inferred albedos, there is an indirect effect due to blended companion stars. When a blend is unrecognized, the planet-hosting star will be inferred to be more luminous, hence larger and more massive than it actually is (assuming main sequence stars). The planet's orbital semi-major axis will be inferred to be too large, and the planet's radius too large. The albedo, proportional to $(a/R_p)^2$ from Equation \ref{eq:refldepth}, will be underestimated, because the effect on $R_p$ will dominate. We simulated this effect for our shortened 1 to 2 $R_{\earth}$ sample, based on the results from \citet{hirsch}. The authors found that one third of planet-hosting stars have companions. We used a Monte-Carlo procedure of assigning companions randomly to one third of our sample, drawing the magnitude difference also randomly from the list of  \citet{hirsch}.
We calculated the effect on each inferred planetary radius and semi-major axis, and thus on the average albedo of the group. We find that a companion fraction of one-third would cause us to underestimate the albedo by 4\%. (Thus an albedo of 0.10 would be inferred as 0.096). However, adopting a one-third companion fraction somewhat exaggerates the effect: The compilation of high-resolution imaging by \citet{furlan} shows companions associated with only 10 of 40 stars in our shortened 1 to 2 $R_{\earth}$ group, less than the one-third from \citet{hirsch}.  Therefore this source of bias is not important for our results, and that is also true for our other planet groups, which have similar companion rates.  Furthermore, we conclude that dilution does not have a significant effect on which radius bin the candidates fall in, because the radius correction factors in \citet{furlan} suggest that only 1 candidate (KOI-2797.01) in the 1 to 2 \rearth~bin would move up to the larger 2 to 4 \rearth~group, while 2 of the 2 to 4 \rearth~candidates would move up to the larger 4 to 6 \rearth~group, and 3 members of the 4 to 6 \rearth~group are over 6 \rearth~by only a few percent.

\subsection{Eccentric Eclipsing Binaries}
Eccentric eclipsing binaries lurking in our sample could weaken the average secondary eclipse measured.  We believe that our inspection of individual candidates in Section 3.2, combined with the use of the false positive probabilities from \citet{fpp}, eliminates significant contamination of our sample by conventional eclipsing binaries.  There remains, however, the possibility that unusual eclipsing binaries, such as those with highly eccentric orbits, may still be present in the sample.  The eccentric binary might have a weak secondary eclipse that occurs at a phase much different than 0.5, or the eccentricity may be so high that, due to viewing geometry, only the primary eclipse or the secondary eclipse occurs, but not both.  In either case, it would introduce a flat light curve where we are expecting a secondary eclipse for a planet.  We tested the effect of possible eccentric eclipsing binaries as false positives contaminating our sample.  \citet{santerne} and references therein report that less than 1\% of eclipsing binaries with periods under 10 days have significant eccentricity.  This is the period regime of our candidate sample.  We fit a model to stacked eclipses of each individual candidate to find the candidate's eclipse depth and calculated the weight of the candidate in the overall stacked eclipse.  We then used those weights and candidate eclipse depths to determine a weighted average eclipse depth for the group.  We considered the worst case scenario, where all the candidates in the sample are eclipsing binaries, with 1\% being eccentric \citep{santerne}.  We randomly drew $\sim$ 1\% of the individual eclipse depths and set them to zero to represent an absence of the eclipse at phase 0.5.  We repeated this 100,000 times and found that the resulting average depth varies from the original average depth by at most a few hundredths of a ppm, so we conclude that eccentric eclipsing binaries are not a significant source of concern within our sample.  

\subsection{Half-Period Eclipsing Binaries}
An eclipsing binary consisting of two similar stars would result in a light curve with a primary eclipse and secondary eclipse that have similar depths.  This mimics a planet at half the period of the binary, resulting in a flat light curve at phase 0.5 when phase-folding to the planet candidate period.   The false positive probability calculations of \citet{fpp} take into account this scenario.  Only a few of our original candidates show greater than a 1\% chance of being an eclipsing binary false positive for this scenario, so we do not believe this could be a significant contributor to the low albedos we find in Section 6.

\subsection{Eccentric Planets}
\begin{figure}
\epsscale{1.0}
\plotone{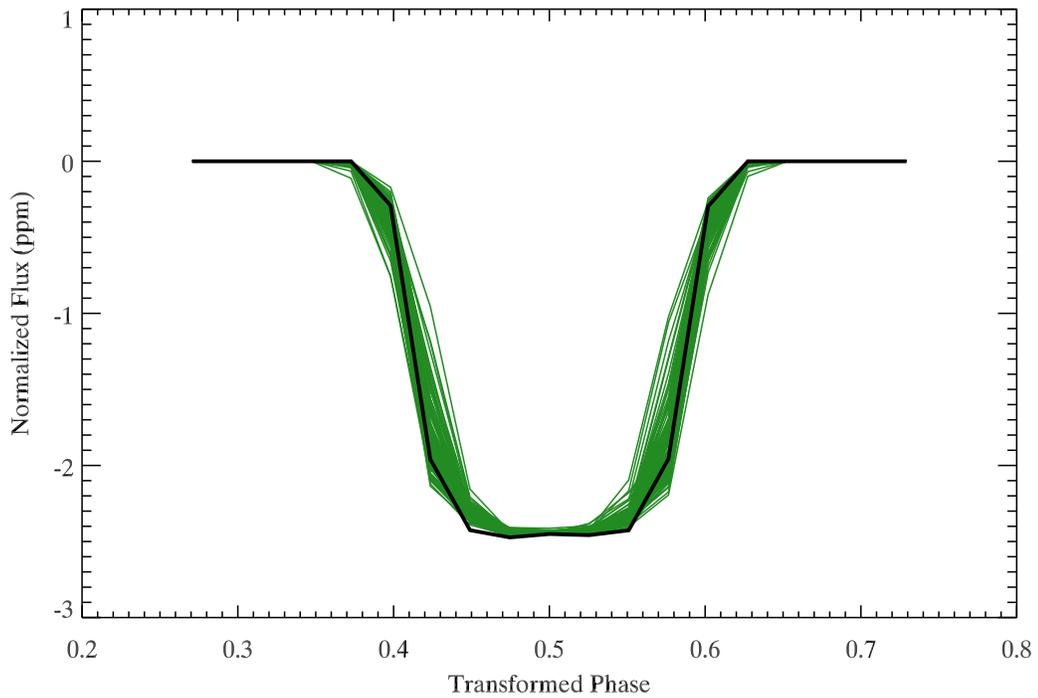}
%\vspace{1.0 in}
\caption{Shown are the averaged model curves for 100 trials of the super-Earth group, in green, assuming eccentricities for the planets drawn from a Rayleigh distribution.  In black is the averaged model curve assuming zero eccentricity for all planets.}
\label{fig:eccplanet}
\end{figure}

We also considered our adoption of circular orbits of the planets.  If any of the planet candidates in our sample are eccentric, the secondary eclipse is displaced from phase 0.5.  To test how this affects our results, we assigned each object in the super-Earth group (excluding Kepler 10b) a random eccentricity, drawing from a Rayleigh distribution with scale parameter $\sigma_e$ = 0.017.  \citet{eccplanets} found that small planets ($<$ 2.5 \rearth) in multi-planet systems have eccentricities with this distribution, while larger planets have smaller eccentricities, with $\sigma_e$ = 0.008.  Because planets in multi-planet systems can have non-zero eccentricities due to mutual perturbations, this is a worst-case scenario for our sample, which contains many single-planet systems.  After assigning a random eccentricity and a random argument of pericenter (uniformly distributed from 0 to 2$\pi$), we calculated a thermal plus reflected light model for each individual eclipse for every candidate, assuming $f = 1/4$ and $A_{g}$ = 0.108, which is the average albedo we found for the group (see Section 6.1).  We then averaged the model eclipse curves just as the data are averaged.  We repeated this for a total of 100 trials, and the resulting averaged model curves are shown in green in Figure \ref{fig:eccplanet}.  Overplotted in black is the model curve using zero eccentricity for all of the candidates.  The 100 averaged model curves have a mean depth of 2.449 ppm, and the shallowest averaged model curve has a depth of 2.412 ppm, versus the depth for the zero eccentricity of 2.473 ppm.  The difference is similar to the error bars on the measured depth of the averaged data, which means that this effect does not drastically change the measured albedo for the group. 

We mention one other interpretation of our results, namely that high eccentricities could be common among our planet groups. Significant eccentricities in our sample would cause us to miss many of the secondary eclipses, and would produce low apparent albedos.  \citet{hat11} measured the eccentricity of HAT-P-11b to be 0.26 based on detection of the secondary eclipse in Kepler data.  They find a geometric albedo of 0.39 $\pm$ 0.07, significantly higher than we infer for our 4 to 6 \rearth~group.  In a statistical study, \citet{xie} infer a mean eccentricity for 0.32 for Kepler's single-planet systems, but there seems to be two populations of single planets.  Most are in low-eccentricity orbits, while 16\%-36\% of single planets form a population with much higher average eccentricities, which inflates the overall mean.  Furthermore, the eccentricity trends lower with decreasing orbital period, albeit with large uncertainty (Xie, private communication).  Given the strong tendency toward tidal circularization for orbital periods typical of the planets in our sample, we believe that significant orbital eccentricities is a radical and unlikely interpretation of our results, but we mention it for completeness.

\section{Results and Discussion} % section 6

\subsection{Super-Earths}
\subsubsection{Excluding Kepler-10b}
As in Paper I, we average the 1-2 \rearth~group without Kepler-10b, because Kepler-10 dominates the average.  Kepler-10 is a bright star, and the weighting is done by the photometric errors.  Kepler-10b also has an unusually deep secondary eclipse, requiring either a high albedo or some emission source beyond the blackbody emission for a planet in equilibrium with the incoming stellar light.  To determine the fitted depth of the averaged data, we generate a model for the eclipse.  We create a model point that corresponds to each data point, using the \citet{agol} model, scaled so that the continuum is 0 and full eclipse is -1, and averaged over the 30 min exposure for the corresponding data point.  We average the model points with the same weights as the corresponding data points.  We then use a simple 2-parameter Markov Chain Monte Carlo (MCMC) procedure with 500,000 steps, as in Paper I, fitting the equation $F_b + x\delta$, where $F_b$ adjusts the continuum level, $x$ is the model, and $\delta$ is the depth of the eclipse.  We choose the MCMC approach in order to generate error distributions and capture possible correlations between the eclipse depth and out-of-eclipse offset value.

\begin{figure}
%\vspace{0.7 in}
%\epsscale{0.8}
\epsscale{1.0}
\plotone{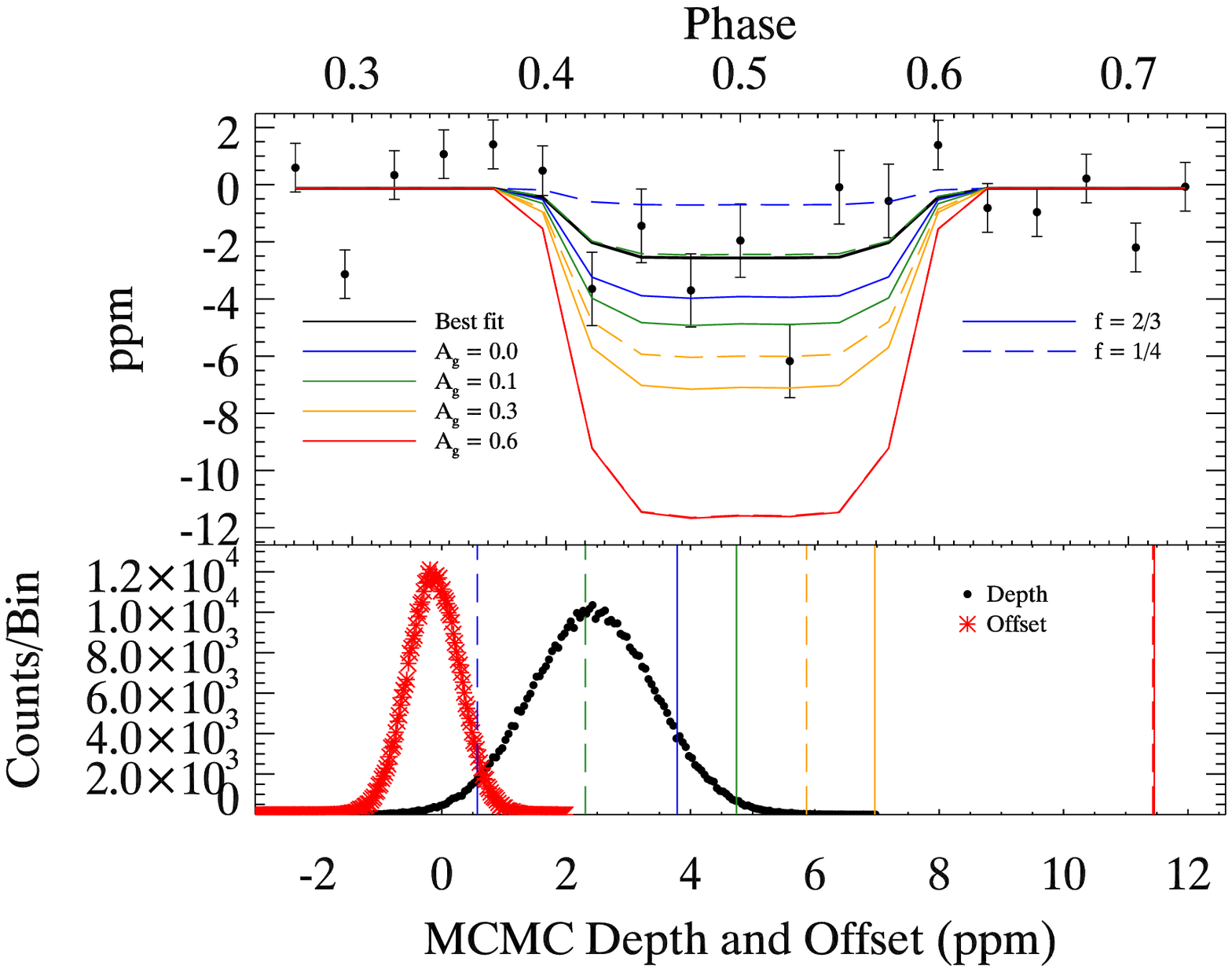}
\vspace{1.0 in}
\caption{Upper panel:  The averaged light curve, centered on secondary eclipse, for the super-Earth group using the 2015 February catalog, excluding Kepler-10b.  The binned data are shown as points.  The error bars are the propagated photometric errors.  The best fit curve is the solid black line.  Overplotted are the reflected light plus thermal emission models for $A_g = (2/3)*A_B =$ 0.0 (blue), 0.1 (green), 0.3 (orange), and 0.6 (red), with the re-radiation factor $f =$ 1/4 (dashed) and 2/3 (solid).   Lower panel:  The distributions for the two parameters of the MCMC run, with the depths from the reflected light plus thermal emission from the upper panel plotted as vertical lines.  The two fitted parameters are eclipse depth (2.44 $\pm$ 0.99 ppm) and continuum offset from zero (-0.12 $\pm$ 0.42 ppm).}
\label{fig:plot1}
\end{figure}

\begin{figure}
%\epsscale{1.1}
\epsscale{1.0}
\plotone{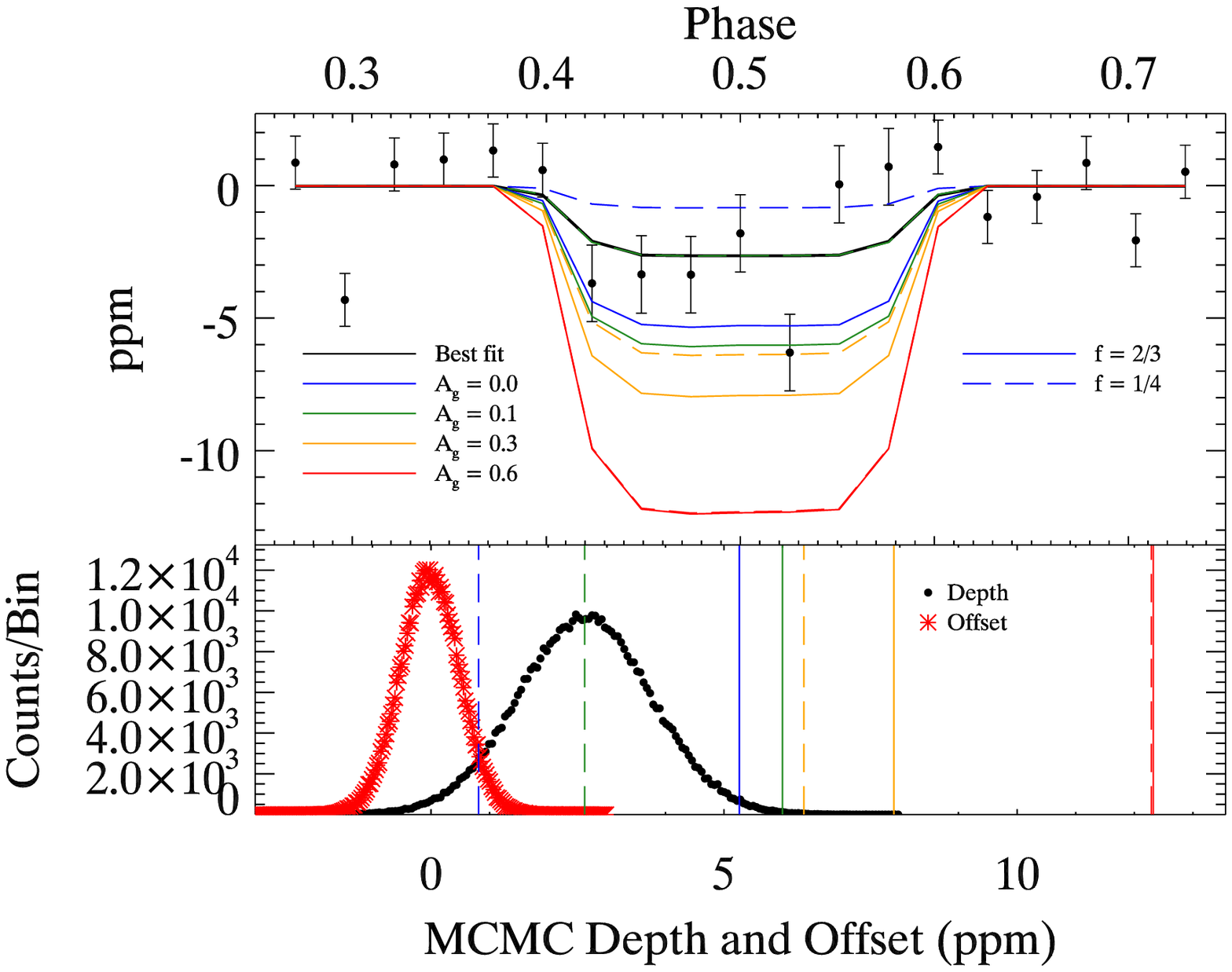}
\vspace{1.0 in}
\caption{Upper panel:  The averaged light curve, centered on secondary eclipse, for the shortened super-Earth group using the finalized catalog and false positive probabilities, excluding Kepler-10b.  The binned data are shown as points.  The error bars are the propagated photometric errors.  The best fit curve is the solid black line.  Overplotted are the reflected light plus thermal emission models for $A_g = (2/3)*A_B =$ 0.0 (blue), 0.1 (green), 0.3 (orange), and 0.6 (red), with the re-radiation factor $f =$ 1/4 (dashed) and 2/3 (solid).   Lower panel:  The distributions for the two parameters of the MCMC run, with the depths from the reflected light plus thermal emission from the upper panel plotted as vertical lines.  The two fitted parameters are eclipse depth (2.6 $\pm$ 1.1 ppm) and continuum offset from zero (-0.02 $\pm$ 0.50 ppm).}
\label{fig:plot1a}
\end{figure}

The result for the 2015 February list of super-Earths is shown in Figure \ref{fig:plot1}.  This group contains 55 candidates and 79,678 individual eclipses.  The weighted average depth is 2.44 $\pm$ 0.99 ppm, which corresponds to a geometric albedo $A_g$ = 0.11 $\pm$ 0.06 in the case of complete redistribution of heat.  In the case of instantaneous re-radiation, the measured depth corresponds to an unphysical, negative value for albedo, but we can set an upper, 3$\sigma$ limit of $A_g <$ 0.17.  The shortened group, removing false positives identified in the finalized catalogs and candidates with high false positive probability, gives similar results, shown in Figure \ref{fig:plot1a}.   The shortened group contains 39 candidates and 50,805 eclipses.  The weighted average depth is 2.63 $^{+1.13}_{-1.14}$ ppm, but the weighted average \rpasq~increases as well, from 19.06 to 20.48, and so the change in eclipse depth is not sufficient to change the values of the albedos significantly.  We know that the 2015 February list contained one definite false positive, KOI 1662.02, which was removed from the shortened list.  Including KOI 1662.02 in the average was adding a flat curve at secondary eclipse.  Since removing it did not appreciably change the group's average albedo, this suggests that most of the candidates are indeed dark, rather than the average being pulled down by false positives.

\subsubsection{Including Kepler-10b}
\begin{figure}
%\vspace{0.7 in}
\epsscale{1.0}
\plotone{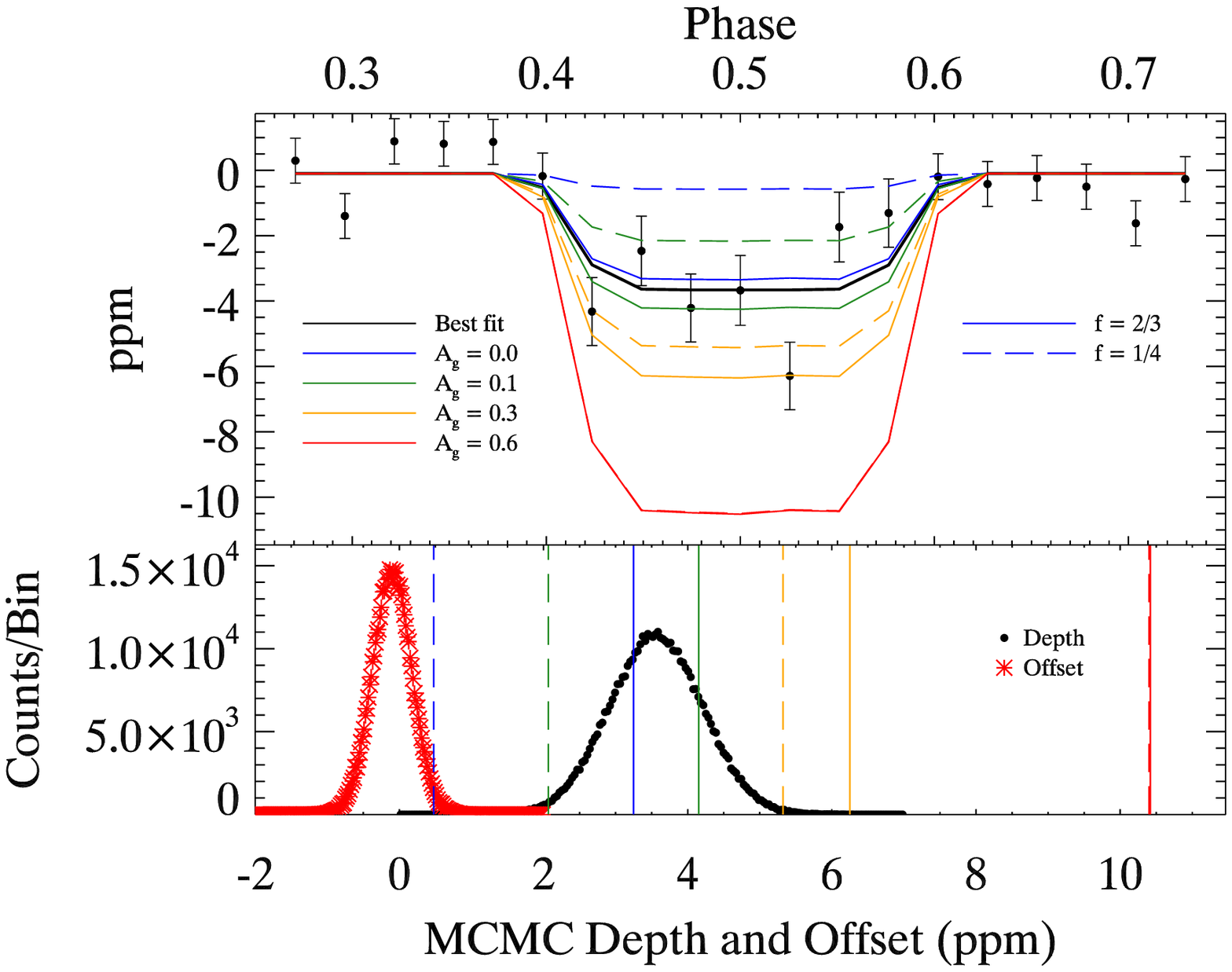}
\vspace{1.0 in}
\caption{Upper panel:  The averaged light curve, centered on secondary eclipse, for the super-Earth group with the 2015 February catalog, including Kepler-10b.  The binned data are shown as points.  The error bars are the propagated photometric errors.  The best fit curve is the solid black line.  Overplotted are the reflected light plus thermal emission models for $A_g = (2/3)*A_B =$ 0.0 (blue), 0.1 (green), 0.3 (orange), and 0.6 (red), with the re-radiation factor $f =$ 1/4 (dashed) and 2/3 (solid).   Lower panel:  The distributions for the two parameters of the MCMC run, with the depths from the reflected light plus thermal emission from the upper panel plotted as vertical lines.  The two fitted parameters are eclipse depth (3.56 $\pm$ 0.65 ppm) and continuum offset from zero (-0.10 $\pm$ 0.27 ppm).}
\label{fig:plot2}
\end{figure}

\begin{figure}
\epsscale{1.0}
\plotone{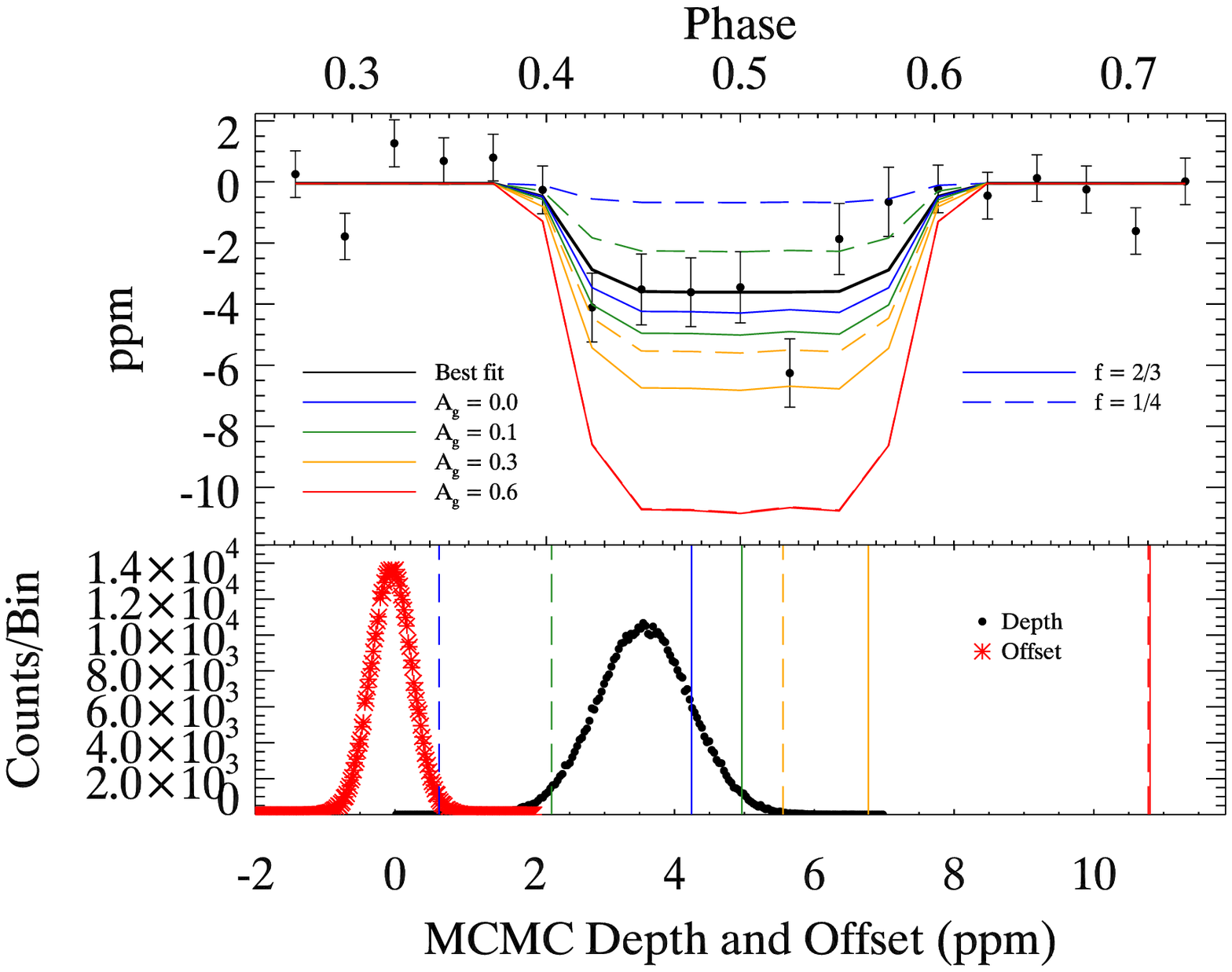}
\vspace{1.0 in}
\caption{Upper panel:  The averaged light curve, centered on secondary eclipse, for the shortened super-Earth group with the finalized catalog and false positive probabilities, including Kepler-10b.  The binned data are shown as points.  The error bars are the propagated photometric errors.  The best fit curve is the solid black line.  Overplotted are the reflected light plus thermal emission models for $A_g = (2/3)*A_B =$ 0.0 (blue), 0.1 (green), 0.3 (orange), and 0.6 (red), with the re-radiation factor $f =$ 1/4 (dashed) and 2/3 (solid).   Lower panel:  The distributions for the two parameters of the MCMC run, with the depths from the reflected light plus thermal emission from the upper panel plotted as vertical lines.  The two fitted parameters are eclipse depth (3.56 $\pm$ 0.67 ppm) and continuum offset from zero (-0.05 $\pm$ 0.29 ppm).}
\label{fig:plot2a}
\end{figure}

Adding Kepler-10b to the group, not surprisingly, increases the average albedo.  The result for the 2015 February group is shown in Figure \ref{fig:plot2}, with 56 candidates and 80,492 individual eclipses.  The weighted average eclipse depth is 3.56 $\pm$ 0.65 ppm, corresponding to an average geometric albedo of $A_g$ = 0.19 $\pm$ 0.04 in the case of full heat redistribution, or $A_g$ = 0.04 $\pm$ 0.07 for instantaneous re-radiation.  The shortened list result, consisting of 40 candidates and 51,597 eclipses, is shown in Figure \ref{fig:plot2a}, with the same eclipse depth of 3.56 ppm, but slightly larger error bars of $\pm$ 0.67 ppm.  This results in the same geometric albedo as the 2015 February group for the case of full heat redistribution, but a slightly (though not significantly) lower $A_g$ = 0.02$^{+0.07}_{-0.08}$  for the case of instantaneous re-radiation.

\subsubsection{Using Individual Candidate Depths}
\begin{figure}
\epsscale{1.0}
\plotone{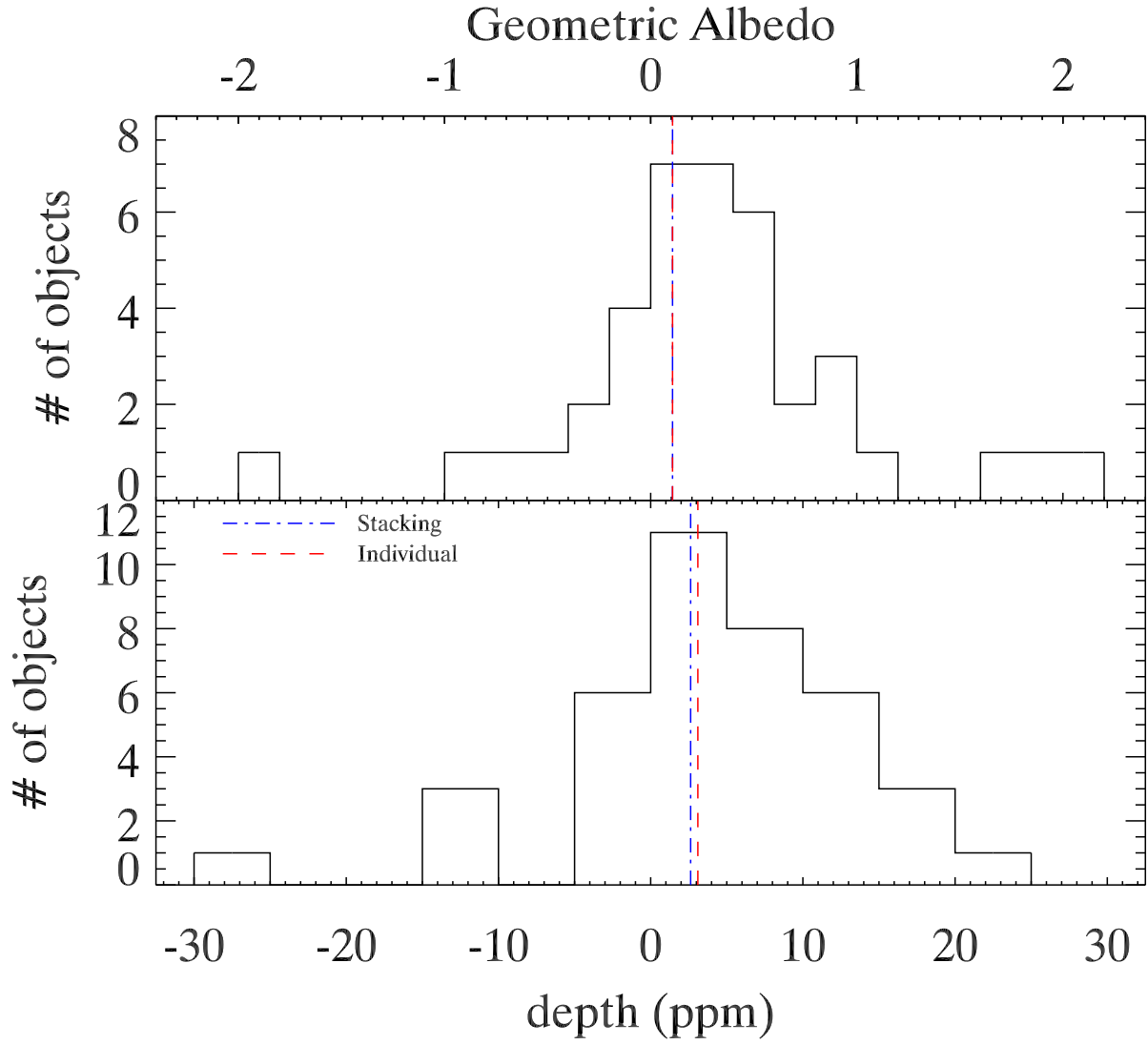}
\vspace{1.0in}
\caption{Upper panel:  Shown is the distribution of geometric albedos determined, assuming only reflected light, from the eclipse depths found for the individual candidates in the shortened 1 to 2 \rearth~group.  Lower Panel:  Shown is the distribution of eclipse depths found for the individual candidates in the same group.  The average values from the stacked average are plotted in the blue dash-dot line, while the average values from the individual candidate depths are plotted in the red dashed line.}
\label{fig:depthhist}
\end{figure}

We also consider the eclipse depths that we fit to each individual candidate while screening for significant, non-planetary eclipses.  These depths are fit using a simple linear regression, rather than the MCMC used on the stacked data of many candidates, so the uncertainties are likely underestimated.  The histogram of depths, for the shortened group without Kepler-10b, is shown in the bottom panel of Figure \ref{fig:depthhist}.   We used Equation \ref{eq:refldepth} to approximate the geometric albedo from the depths, and the histogram of albedos is shown in the top panel of Figure \ref{fig:depthhist}.  Weighting the depths by their uncertainties gives an average depth of 3.10 ppm, vs 2.63 ppm from the stacked average.  Weighting the albedos by their uncertainties gives an average albedo of 0.11, which matches the 0.11 found from the stacked average.  The stacked average, however, accounts for a thermal contribution.  Neglecting thermal emission gives an albedo of 0.13 from the stacked average.  The values from the stacked average are plotted as a blue dash-dot line in Figure \ref{fig:depthhist}, while the values from the individual candidate depths are plotted as a red dashed line.  This suggests we might be slightly overestimating the albedo using the stacked average.

\subsection{Mini-Neptunes}
\begin{figure}
%\vspace{0.7 in}
\epsscale{1.0}
\plotone{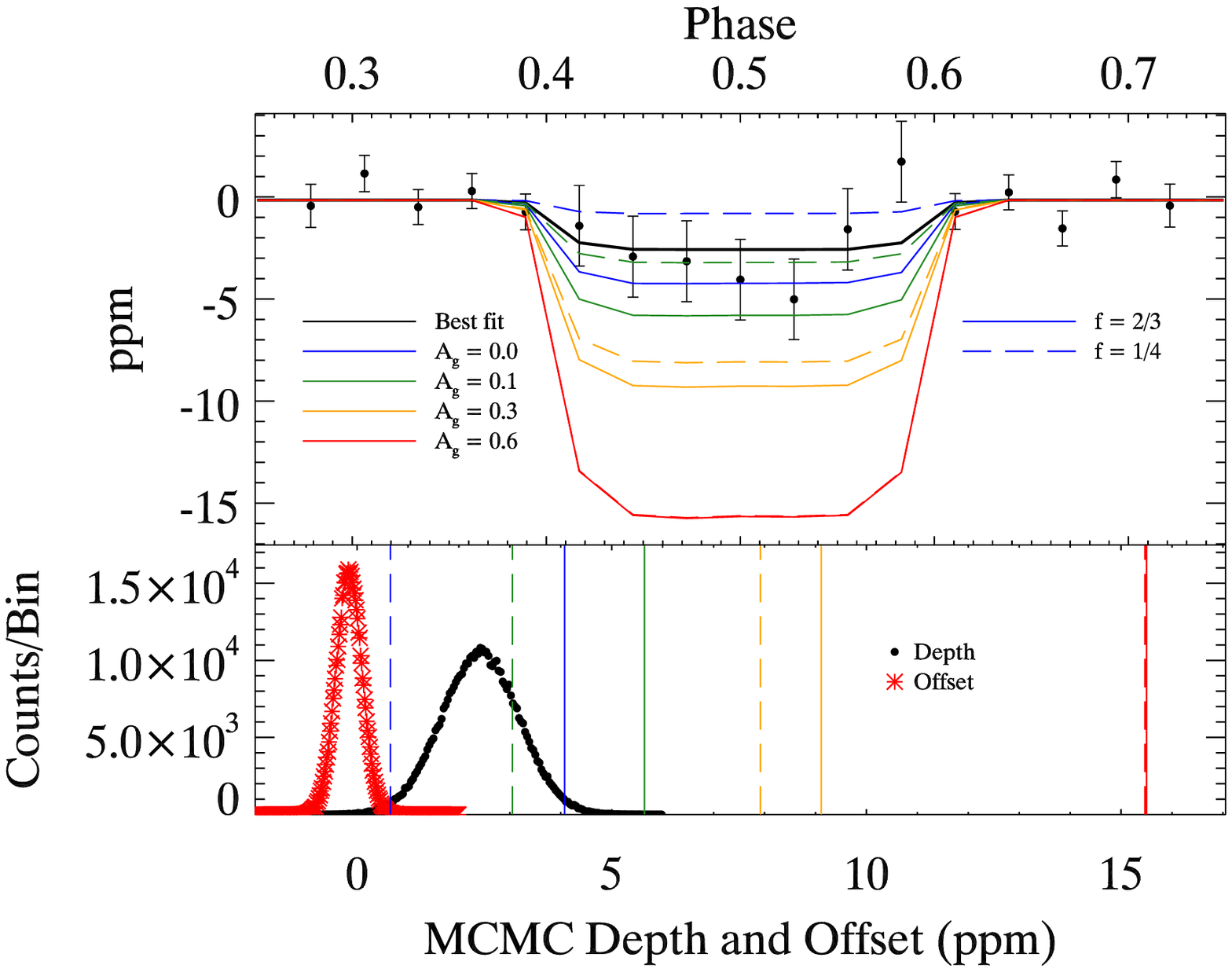}
\vspace{1.0 in}
\caption{Upper panel:  The averaged light curve, centered on secondary eclipse, for the mini-Neptune group with the 2015 February catalog.  The binned data are shown as points.  The error bars are the propagated photometric errors.  The best fit curve is the solid black line.  Overplotted are the reflected light plus thermal emission models for $A_g = (2/3)*A_B =$ 0.0 (blue), 0.1 (green), 0.3 (orange), and 0.6 (red), with the re-radiation factor $f =$ 1/4 (dashed) and 2/3 (solid).   Lower panel:  The distributions for the two parameters of the MCMC run, with the depths from the reflected light plus thermal emission from the upper panel plotted as vertical lines.  The two fitted parameters are eclipse depth (2.42 $\pm$ 0.76 ppm) and continuum offset from zero (-0.16 $\pm$ 0.25 ppm).}
\label{fig:plot3}
\end{figure}

\begin{figure}
\epsscale{1.0}
\plotone{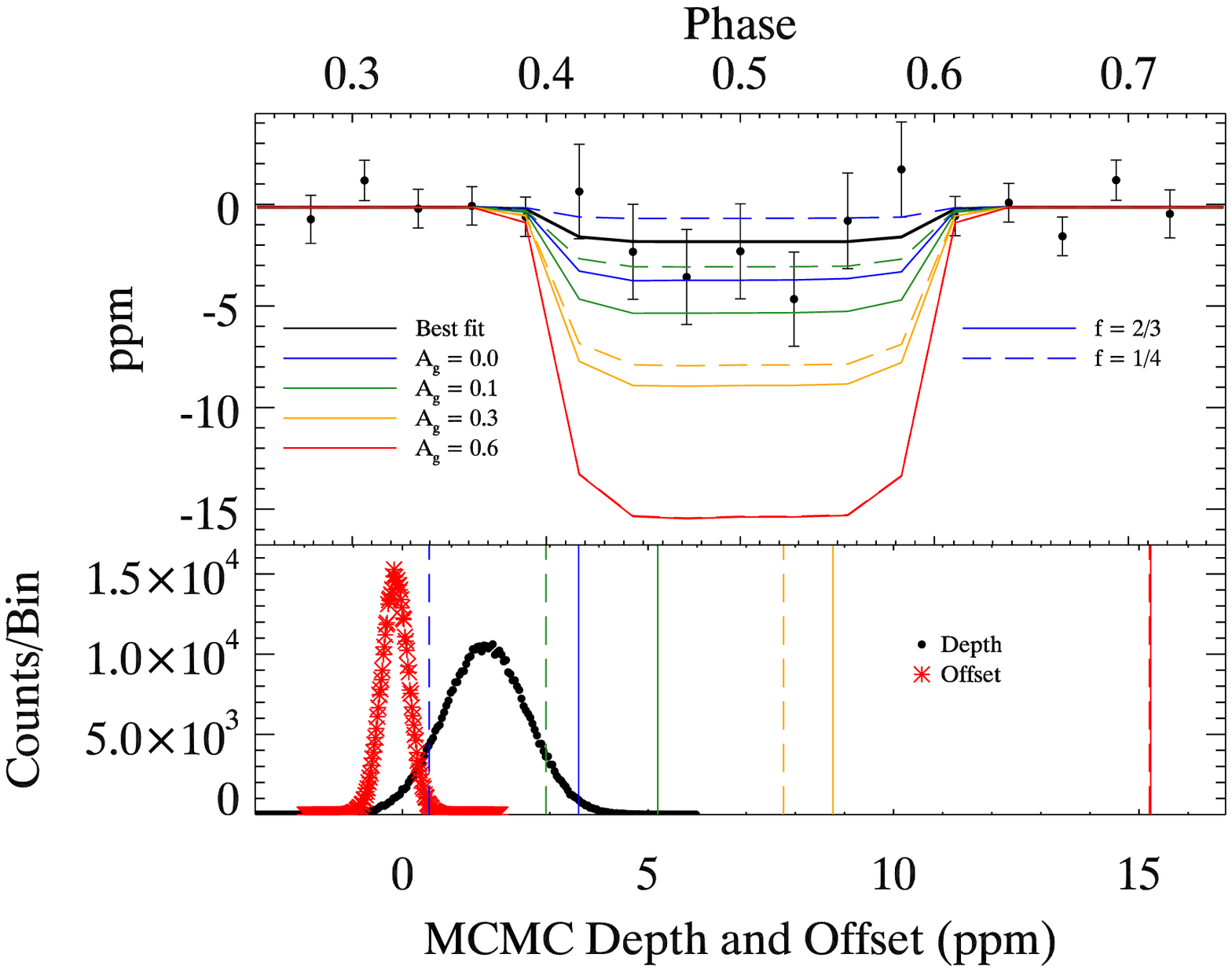}
\vspace{1.0 in}
\caption{Upper panel:  The averaged light curve, centered on secondary eclipse, for the shortened mini-Neptune group with the finalized catalog and false positive probabilities.  The binned data are shown as points.  The error bars are the propagated photometric errors.  The best fit curve is the solid black line.  Overplotted are the reflected light plus thermal emission models for $A_g = (2/3)*A_B =$ 0.0 (blue), 0.1 (green), 0.3 (orange), and 0.6 (red), with the re-radiation factor $f =$ 1/4 (dashed) and 2/3 (solid).   Lower panel:  The distributions for the two parameters of the MCMC run, with the depths from the reflected light plus thermal emission from the upper panel plotted as vertical lines.  The two fitted parameters are eclipse depth (1.69 $\pm$ 0.85 ppm) and continuum offset from zero (-0.15 $\pm$ 0.27 ppm).}
\label{fig:plot3a}
\end{figure}

The average secondary eclipse for the 2015 February list of mini-Neptunes is shown in Figure \ref{fig:plot3}.  This group contains 38 candidates and 22,677 eclipses, with an average secondary eclipse depth of 2.42 $\pm$ 0.76 ppm.  The depth corresponds to $A_g$ = 0.07 $\pm$ 0.03 for full heat redistribution.  For instantaneous re-radiation, as with the super-Earths, we can only set an upper, 3$\sigma$ limit, but this limit is much lower, $A_g < 0.04$, than for the super-Earths.  The shortened list consists of 28 candidates and 13,580 eclipses, with the result shown in Figure \ref{fig:plot3a}.  The average secondary eclipse depth is 1.69 $\pm$ 0.85 ppm, corresponding to $A_g$ = 0.05 $\pm$ 0.04 for full heat redistribution.  Again we can only set an upper, 3$\sigma$ limit of $A_g < $ 0.07 for instantaneous re-radiation.  

\subsection{Super-Neptunes} 
\begin{figure}
%\vspace{0.7 in}
\epsscale{1.0}
\plotone{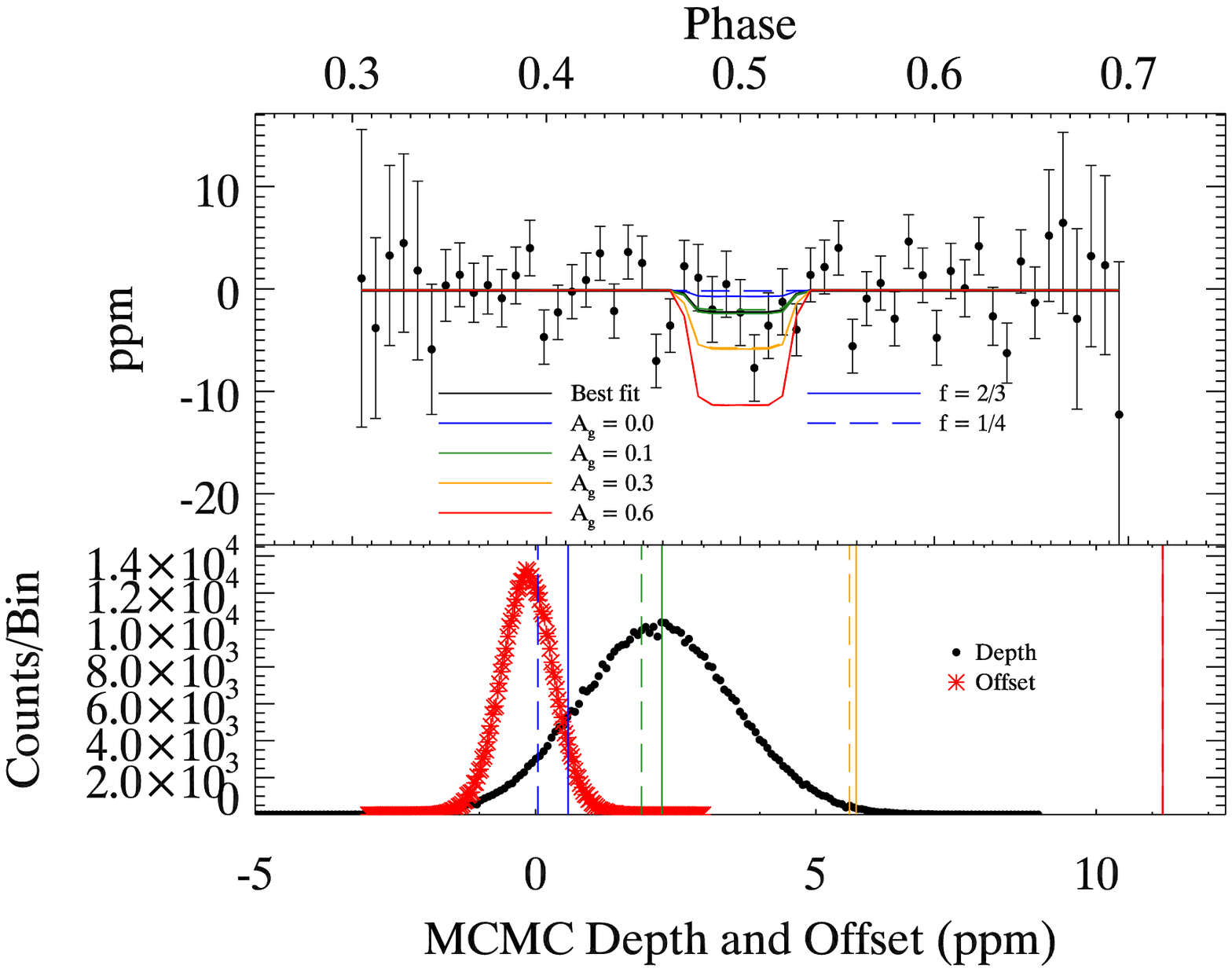}
\vspace{1.0 in}
\caption{Upper panel:  The averaged light curve, centered on secondary eclipse, for the super-Neptune group with the 2015 February catalog.  The binned data are shown as points.  The error bars are the propagated photometric errors.  The best fit curve is the solid black line.  Overplotted are the reflected light plus thermal emission models for $A_g = (2/3)*A_B =$ 0.0 (blue), 0.1 (green), 0.3 (orange), and 0.6 (red), with the re-radiation factor $f =$ 1/4 (dashed) and 2/3 (solid).   Lower panel:  The distributions for the two parameters of the MCMC run, with the depths from the reflected light plus thermal emission from the upper panel plotted as vertical lines.  The two fitted parameters are eclipse depth (2.16 $\pm$ 1.38 ppm) and continuum offset from zero (-0.14 $\pm$ 0.46 ppm).}
\label{fig:plot4}
\end{figure}

\begin{figure}
\epsscale{1.0}
\plotone{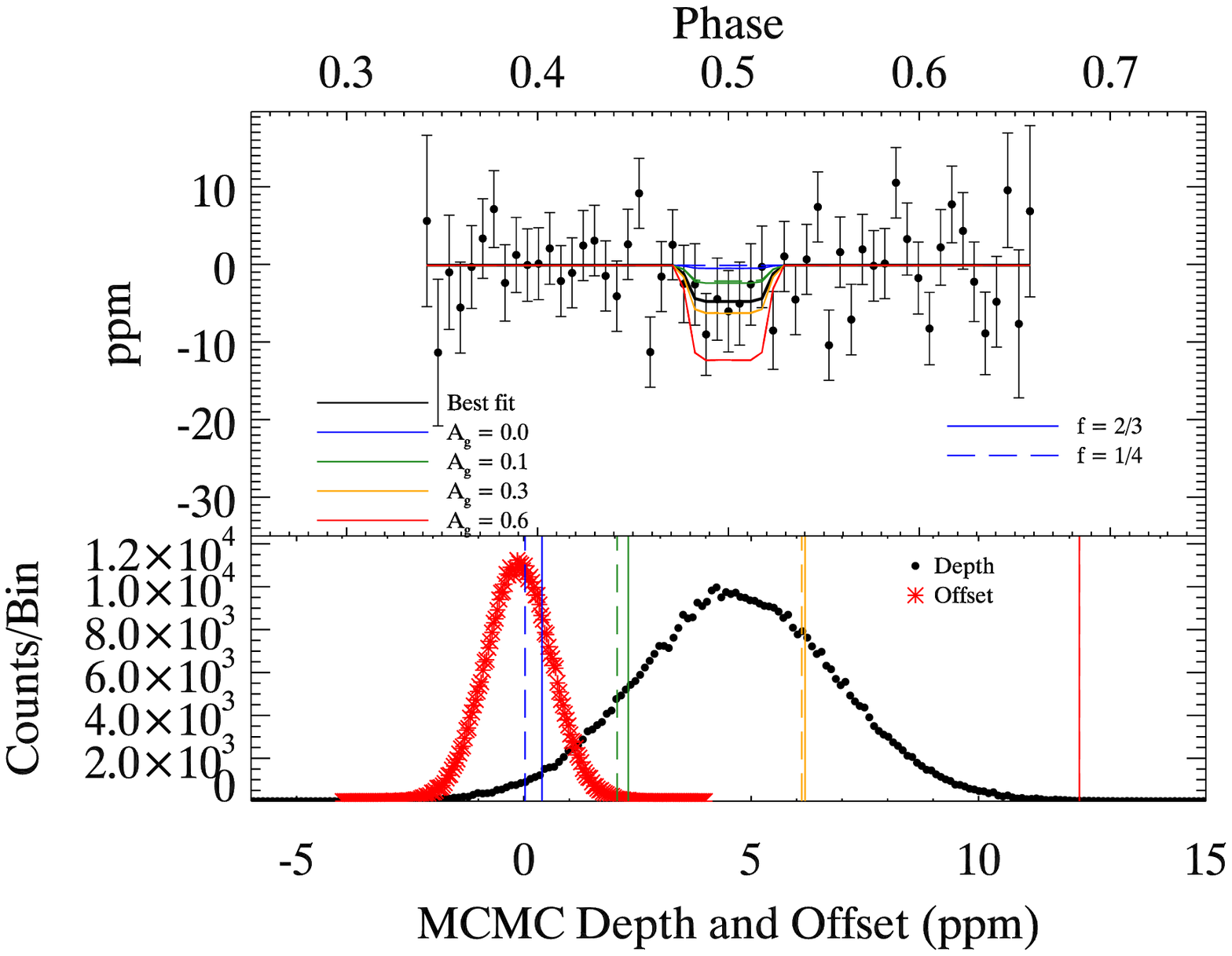}
\vspace{1.0 in}
\caption{Upper panel:  The averaged light curve, centered on secondary eclipse, for the shortened super-Neptune group with the finalized catalog and false positive probabilities.  The binned data are shown as points.  The error bars are the propagated photometric errors.  The best fit curve is the solid black line.  Overplotted are the reflected light plus thermal emission models for $A_g = (2/3)*A_B =$ 0.0 (blue), 0.1 (green), 0.3 (orange), and 0.6 (red), with the re-radiation factor $f =$ 1/4 (dashed) and 2/3 (solid).   Lower panel:  The distributions for the two parameters of the MCMC run, with the depths from the reflected light plus thermal emission from the upper panel plotted as vertical lines.  The two fitted parameters are eclipse depth (4.67$^{+2.18}_{-2.15}$ ppm) and continuum offset from zero (-0.12 $\pm$ 0.73 ppm).}
\label{fig:plot4a}
\end{figure}

Figure \ref{fig:plot4} shows the result for the 2015 February group of super-Neptunes, which contains 16 candidates and 4,572 eclipses.  The average secondary eclipse depth is 2.16$^{+1.37}_{-1.38}$ ppm.  This depth corresponds to a geometric albedo $A_g$ = 0.12 $\pm$ 0.08 for full heat redistribution and $A_g$ = 0.09 $\pm$ 0.08 for instantaneous re-radiation.  The shortened group contains 12 candidates and 3,316 eclipses, and the result is shown in Figure \ref{fig:plot4a}.  The average secondary eclipse depth is 4.67$^{+2.18}_{-2.15}$ ppm, which corresponds to $A_g$ = 0.23 $\pm$ 0.11 for full redistribution of heat and $A_g$ = 0.21 $\pm$ 0.11 for instantaneous re-radiation.  

\subsection{Comparison to Short Cadence}
\begin{figure}
%\vspace{0.7 in}
\epsscale{1.0}
\plotone{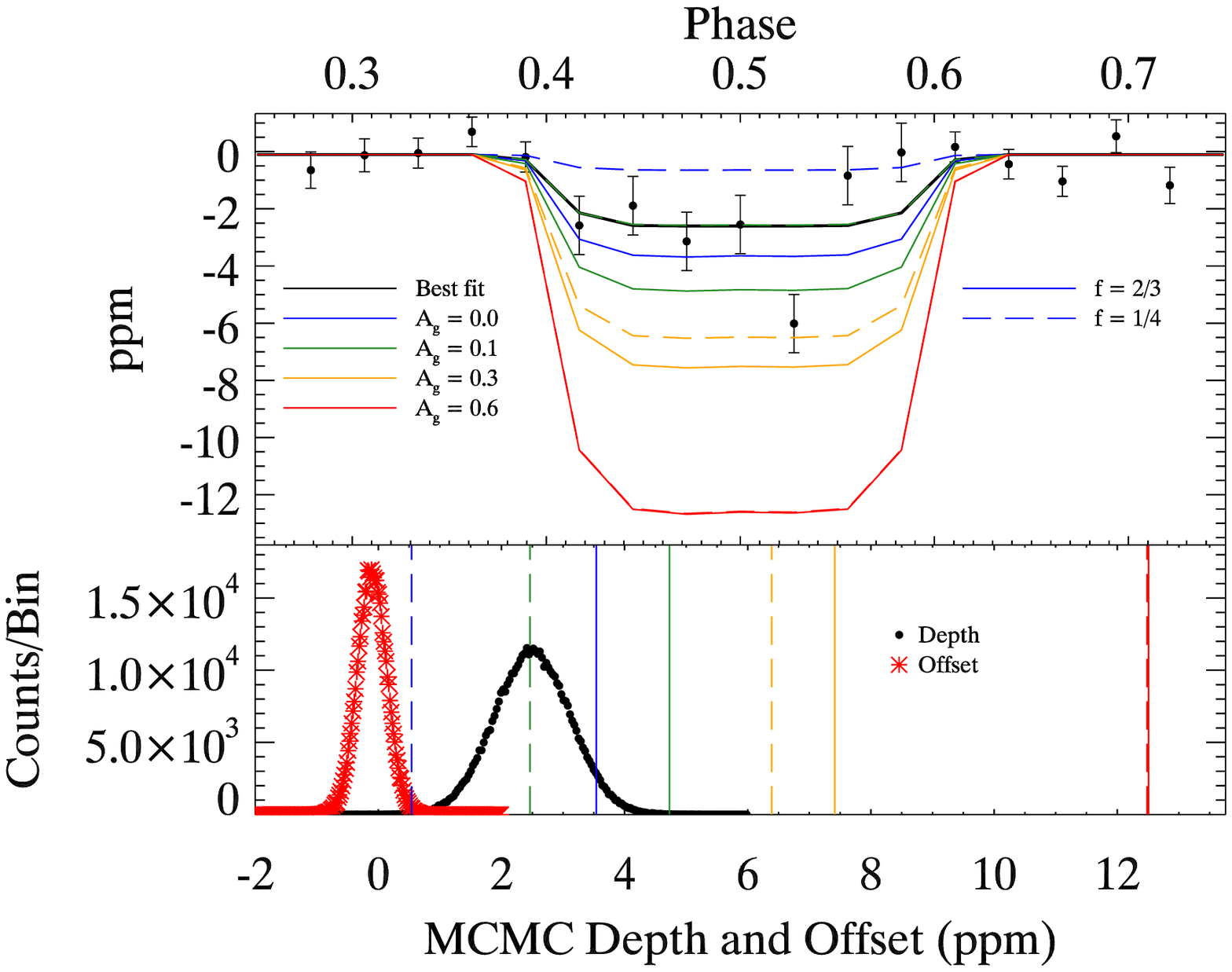}
\vspace{1.0 in}
\caption{Upper panel:  The averaged light curve, centered on secondary eclipse, for the 1 to 6 R$_{\earth}$ group with the 2015 February catalog, excluding Kepler-10b.  The binned data are shown as points.  The error bars are the propagated photometric errors.  The best fit curve is the solid black line.  Overplotted are the reflected light plus thermal emission models for $A_g = (2/3)*A_B =$ 0.0 (blue), 0.1 (green), 0.3 (orange), and 0.6 (red), with the re-radiation factor $f =$ 1/4 (dashed) and 2/3 (solid).   Lower panel:  The distributions for the two parameters of the MCMC run, with the depths from the reflected light plus thermal emission from the upper panel plotted as vertical lines.  The two fitted parameters are eclipse depth (2.50 $\pm$ 0.62 ppm) and continuum offset from zero (-0.10 $\pm$ 0.24 ppm).}
\label{fig:plot5}
\end{figure}

\begin{figure}
\epsscale{1.0}
\plotone{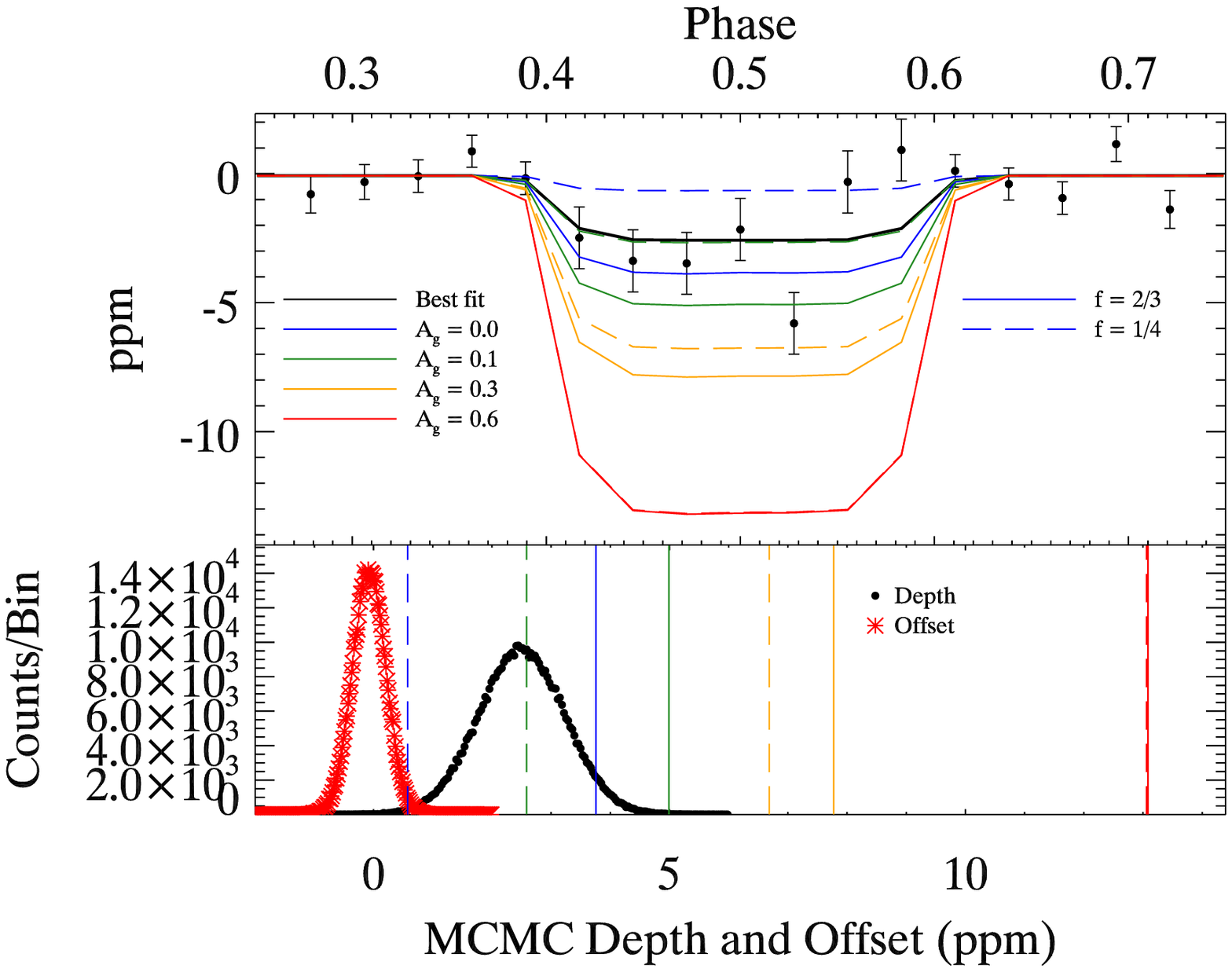}
\vspace{1.0 in}
\caption{Upper panel:  The averaged light curve, centered on secondary eclipse, for the shortened 1 to 6 R$_{\earth}$ group with the finalized catalog and false positive probabilities, excluding Kepler-10b.  The binned data are shown as points.  The error bars are the propagated photometric errors.  The best fit curve is the solid black line.  Overplotted are the reflected light plus thermal emission models for $A_g = (2/3)*A_B =$ 0.0 (blue), 0.1 (green), 0.3 (orange), and 0.6 (red), with the re-radiation factor $f =$ 1/4 (dashed) and 2/3 (solid).   Lower panel:  The distributions for the two parameters of the MCMC run, with the depths from the reflected light plus thermal emission from the upper panel plotted as vertical lines.  The two fitted parameters are eclipse depth (2.50$^{+0.73}_{-0.72}$ ppm) and continuum offset from zero (-0.08 $\pm$ 0.28 ppm).}
\label{fig:plot5a}
\end{figure}

\begin{figure}
%\vspace{0.7 in}
\epsscale{1.0}
\plotone{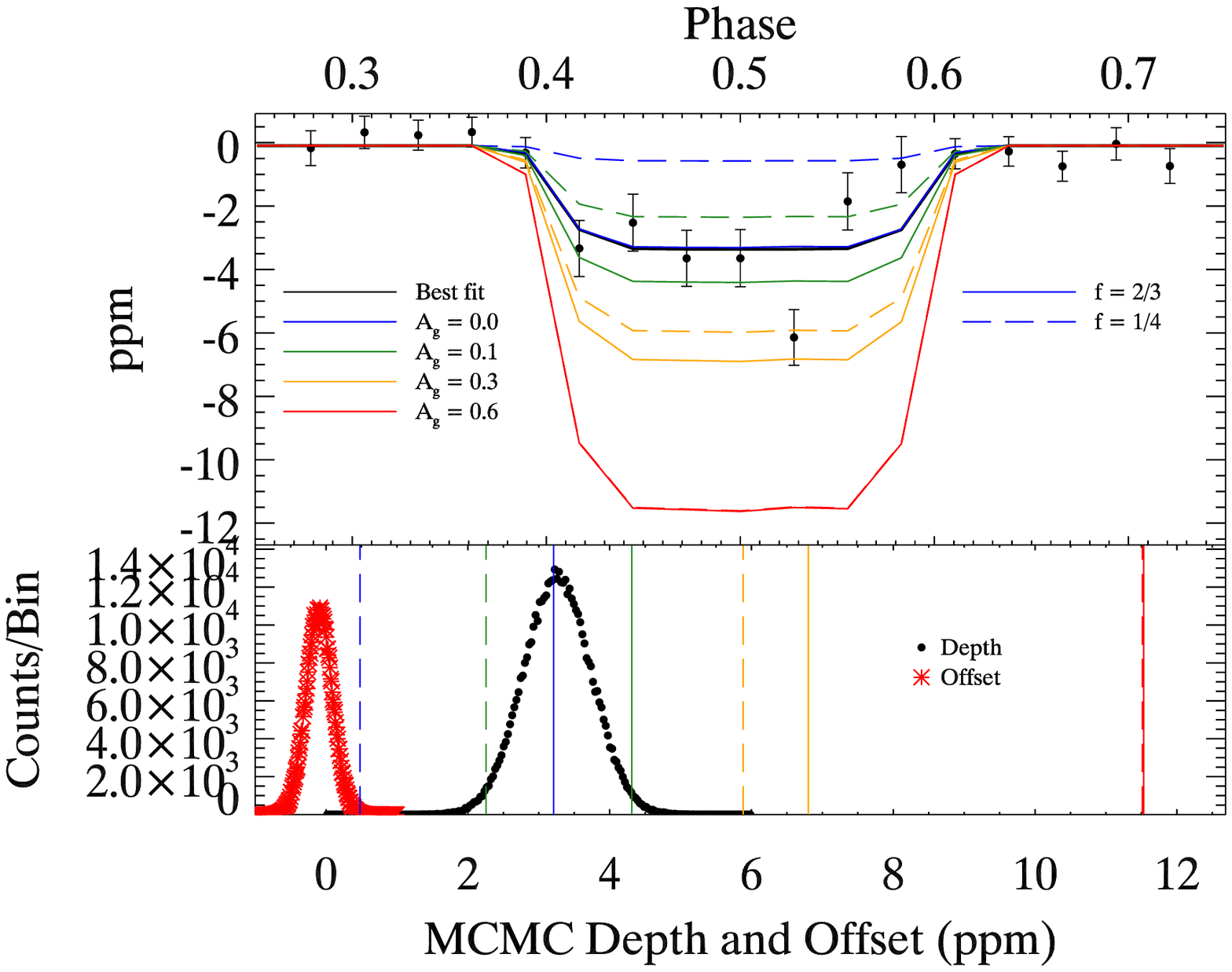}
\vspace{1.0 in}
\caption{Upper panel:  The averaged light curve, centered on secondary eclipse, for the 1 to 6 R$_{\earth}$ group for the 2015 February catalog, including Kepler-10b.  The binned data are shown as points.  The error bars are the propagated photometric errors.  The best fit curve is the solid black line.  Overplotted are the reflected light plus thermal emission models for $A_g = (2/3)*A_B =$ 0.0 (blue), 0.1 (green), 0.3 (orange), and 0.6 (red), with the re-radiation factor $f =$ 1/4 (dashed) and 2/3 (solid).   Lower panel:  The distributions for the two parameters of the MCMC run, with the depths from the reflected light plus thermal emission from the upper panel plotted as vertical lines.  The two fitted parameters are eclipse depth (3.26 $\pm$ 0.48 ppm) and continuum offset from zero (-0.10 $\pm$ 0.18 ppm).}
\label{fig:plot6}
\end{figure}

\begin{figure}
\epsscale{1.0}
\plotone{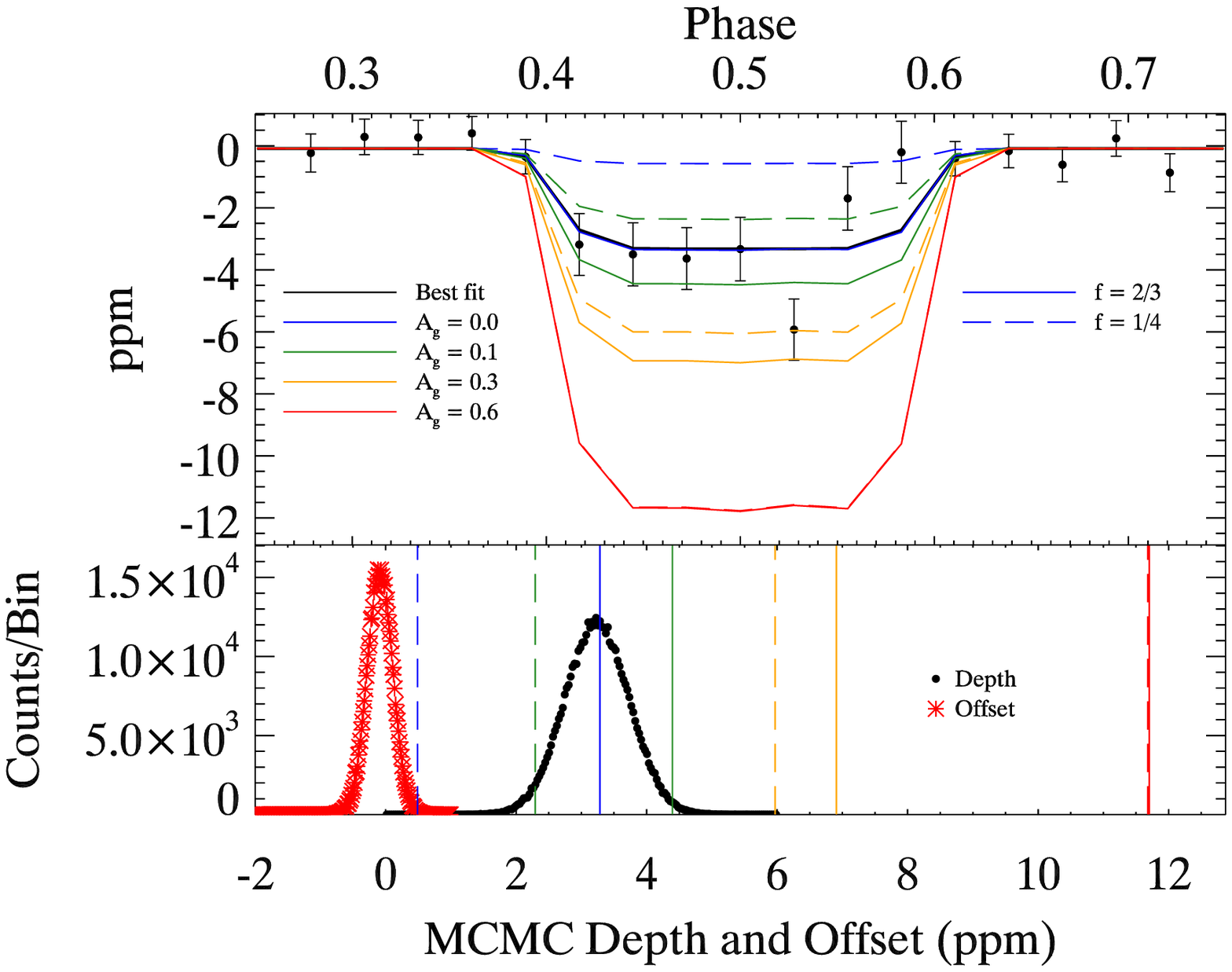}
\vspace{0.5 in}
\caption{Upper panel:  The averaged light curve, centered on secondary eclipse, for the shortened 1 to 6 R$_{\earth}$ group for the finalized catalog and false positive probabilities, including Kepler-10b.  The binned data are shown as points.  The error bars are the propagated photometric errors.  The best fit curve is the solid black line.  Overplotted are the reflected light plus thermal emission models for $A_g = (2/3)*A_B =$ 0.0 (blue), 0.1 (green), 0.3 (orange), and 0.6 (red), with the re-radiation factor $f =$ 1/4 (dashed) and 2/3 (solid).   Lower panel:  The distributions for the two parameters of the MCMC run, with the depths from the reflected light plus thermal emission from the upper panel plotted as vertical lines.  The two fitted parameters are eclipse depth (3.24 $\pm$ 0.49 ppm) and continuum offset from zero (-0.09$^{+0.19}_{-0.20}$ ppm).}
\label{fig:plot6a}
\end{figure}

To compare our results in this paper with those of Paper I, we also take the average of the three radius bins combined, matching the 1-6 \rearth~radius bin of the earlier work.  Figure \ref{fig:plot5} shows the result for the 2015 February group of 1-6 \rearth~candidates, excluding Kepler-10b.  There are 109 candidates for a total of 106,927 eclipses.  The average eclipse depth is 2.50 $\pm$ 0.62 ppm, for $A_g$ = 0.10 $\pm$ 0.03 for full heat redistribution, and $A_g <$ 0.07 for instantaneous re-radiation.  The shortened group result, shown in Figure \ref{fig:plot5a}, is similar, with a depth of 2.50$^{+0.73}_{-0.72}$ ppm, corresponding to $A_g$ = 0.10 $\pm$ 0.04 for full redistribution of heat, and $A_g <$ 0.08 for instantaneous re-radiation.  The shortened group contains 79 candidates, with 67,701 eclipses.  Including Kepler-10b with the 2015 February group, shown in Figure \ref{fig:plot6}, brings the average eclipse depth to 3.26 $\pm$ 0.48 ppm, corresponding to $A_g$ = 0.16 $\pm$ 0.03 for full heat redistribution, and to $A_g$ = 0.01 $\pm$ 0.04 for instantaneous re-radiation.  For the shortened group, shown in Figure \ref{fig:plot6a}, adding Kepler-10b brings the average eclipse depth to 3.24 $\pm$ 0.49 ppm.  This gives $A_g$ = 0.15 $\pm$ 0.03 for full heat redistribution, and $A_g$ = 0.00 $\pm$ 0.05 for instantaneous re-radiation.  The albedos determined from the long cadence are summarized in Table \ref{tab:albresults}.

\begin{figure}
\epsscale{1.3}
\plotone{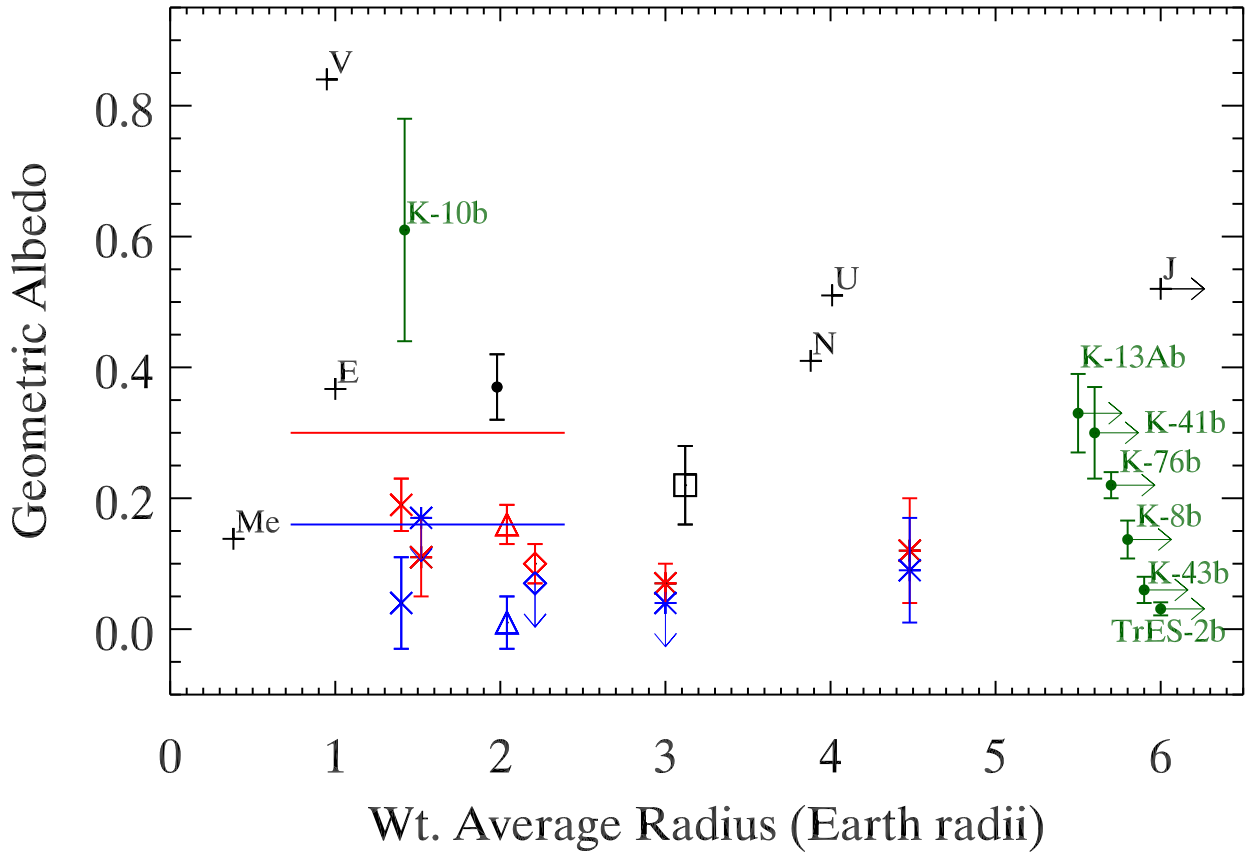}
%\vspace{1.0 in}
\caption{The geometric albedo is plotted versus the weighted average radius for the long cadence data for the super-Earth (excluding Kepler-10b), mini-Neptune, and super-Neptune groups with an asterisk, for the super-Earth group including Kepler-10b with an x (offset by -0.1 in radius for clarity), for the 1 to 6 R$_{\earth}$ group without Kepler-10b with a diamond, and for the 1 to 6 R$_{\earth}$ group with Kepler-10b with a triangle.  Points in red assume a redistribution factor $f = 1/4$, while points in blue assume a redistribution factor of $f = 2/3$.  Also shown in black are the short cadence results for the 1 to 6 R$_{\earth}$ without Kepler-10b (square) and with Kepler-10b (dot).  The error bars show the 1$\sigma$ uncertainty on the albedo.  Points with down arrows instead of error bars (for $f = 2/3$: the super-Earths without Kepler-10b, the mini-Neptunes, and 1 to 6 R$_{\earth}$ without Kepler-10b) represent 3$\sigma$ upper limits.  The radius range and median albedo for the super-Earths in \citet{demory} are shown as the solid red ($f=1/4$) and solid blue ($f=2/3$) lines.  Also included are solar system planets with a black + (Me = Mercury, V = Venus, E = Earth, J = Jupiter, U = Uranus, and N = Neptune).  Plotted with green points is a representative sample of the bright and dark hot Jupiters and Kepler-10b, with values taken from Table \ref{tab:ecltab}.  Right arrows indicate that the measured radius is larger than plotted.  Jupiter and the hot Jupiters are plotted near a radius of 6 to fit within the plot.}
\label{fig:alb}
\end{figure}

For the short cadence data, excluding Kepler-10b, we found $A_g$ = 0.22 $\pm$ 0.06, when only considering reflected light, and the inclusion of thermal emission changed the albedo by less than the uncertainty.  The albedos found from the long cadence data using the 2015 February groups are plotted versus the average group radius in Figure \ref{fig:alb}, along with the short cadence results.  Including Kepler-10b in the short cadence raised the albedo to $A_g$ = 0.37 $\pm$ 0.05.  While the albedos found in this paper are lower than those we found for the short cadence data, excluding Kepler-10b, the difference is only 0.12 $\pm$ 0.07, and so is consistent within 2$\sigma$.   A possible explanation for the lower albedos found here is the greater number of candidates included.  The short cadence analysis only had 31 candidates, plus Kepler-10b, so if any of those candidates are part of the higher-albedo population of candidates like Kepler-10b, it would more strongly influence the mean.  Also note that, in Figure \ref{fig:alb}, the 1-6 \rearth~long cadence group, including Kepler-10b, contains the most data and therefore shows greater separation between the $f=1/4$ and $f=2/3$ thermal models.

There are also eight candidates in our short cadence list that now, with updates to their parameters since Paper I, have radii between 6 and 8 \rearth~(KOIs 356.01 and 1784.01), or have \rpasq $<$ 10 ppm (KOIs 5.01, 299.01, 505.03, 755.01, 1128.01, and 1805.01).  One might expect removing the latter six candidates from the long cadence sample to increase the average albedo, rather than lower it, since their maximum eclipse depths would be smaller than originally expected and thus appear to have a lower albedo.  The albedo does not increase, so perhaps some of these are part of the higher-albedo population or are false positives.  One short cadence candidate, KOI 2678.01, was eliminated from the long cadence group due to a high failure rate of the tests in Section 3.3.1, while another, KOI 2700.01, was not included in the long cadence because of its revised impact parameter.

\subsection{Comparison to \citet{demory}}
The study of 27 long cadence super-Earths by \citet{demory} finds a median geometric albedo $A_g$ = 0.30 after thermal decontamination, assuming a null Bond albedo ($A_B$) and efficient heat redistribution, and a median $A_g$ = 0.16 after thermal decontamination, assuming $A_B$ = 0 and no redistribution of heat (see Figure \ref{fig:alb}).  While we adopt $A_B = (3/2)A_g$ rather than $A_B$ = 0 for thermal removal, this cannot explain our lower mean albedos.  Adopting non-zero Bond albedos means that our equilibrium temperatures overall are lower, therefore weakening the thermal contribution and requiring a higher albedo for a given eclipse depth.   We instead look to several differences in the candidate sample selection to explain our lower albedos.  Only two candidates in our long cadence sample overlap with the sample of \citet{demory}.  Kepler-10b is one of the candidates, and the other is KOI 1169.01, which now has a high false positive probability from \citet{fpp}.  Our long cadence sample is larger, with 56 candidates including Kepler-10b, or 40 for the shortened group with Kepler-10b, and it is comprised of candidates with fainter host stars and larger \rpasq~values.  As with the comparison between our short cadence and our long cadence results, the larger sample size could be a factor, since the error in the mean should decrease with a larger sample size.  Furthermore, the selection cut of \citet{demory} based on a total albedo uncertainty of less than 1.0 may be preferentially selecting high-albedo candidates from the lower signal-to-noise stars.

\subsection{Biases in Our Albedo Due to Averaging}
The average albedo calculated for a group is potentially biased by candidates with larger albedos and larger $(R_p/a)^2$.  By calculating the average eclipse depth, we are essentially averaging a sum of $A_{g}*(R_p/a)^2$ for the group.  If thermal contribution is negligible, the average albedo is given by dividing the average depth by the average $(R_p/a)^2$.  This process does not give the same result as simply averaging the albedos directly.  To test how this affects our results, we assumed a simple uniform distribution for the albedos from 0 to 0.6 and randomly assigned an albedo to each candidate in the Feb 2015 groups.  Using the randomly-assigned albedo and the known \rpasq~value and stellar parameters for each candidate, we calculated the thermal and reflected light model point for each data point that goes into the stacked eclipse, to create a simulated stacked eclipse.  We averaged these thermal and reflected light model points just like the data to produce a weighted average eclipse that mimics the weighted average eclipse from the data, but with a known underlying albedo distribution.  The weight for the simulated point is the photometric noise of the corresponding data point.  We then fit this average eclipse with a depth and offset, just as we fit the real average eclipse data.  We used the same iterative process on this calculated average eclipse to determine the average albedo as the process we use to determine the average albedo with the data.  We can then compare the derived albedo via the simulated eclipse depths to the direct weighted average of the known, assigned albedos.  We estimate the weights of the candidates in the direct weighted average by tracking the weight of each candidate in each phase bin of the averaged real data, as well as the weight of the particular phase bin in the overall fit of the average eclipse.  The weight of each candidate is the sum over all the phase bins of  the product of its weight in the phase bin times the weight of the phase bin in the overall fit.  We performed these calculations 100 times for each group, for each of the 2 values for heat redistribution.  We found that for $f = 1/4$, the derived average albedo varied from the known average albedo by -1$^{+13}_{-10}$\% for the 1-2 \rearth~group (excluding Kepler-10b) and -1$^{+10}_{-8}$\% for the 2-4 \rearth~group, where the error bars define the central 68\% of the trials.  For $f = 2/3$, the derived average albedo varied from the known average albedo by -4$^{+5}_{-7}$\% for the 1-2 \rearth~group (excluding Kepler-10b) and -4$^{+11}_{-10}$\% for the 2-4 \rearth~group.  We conclude that this source of bias is not significant to our analysis.

\subsection{Implications of the Low Albedos}
\begin{figure}
\epsscale{1.0}
\plotone{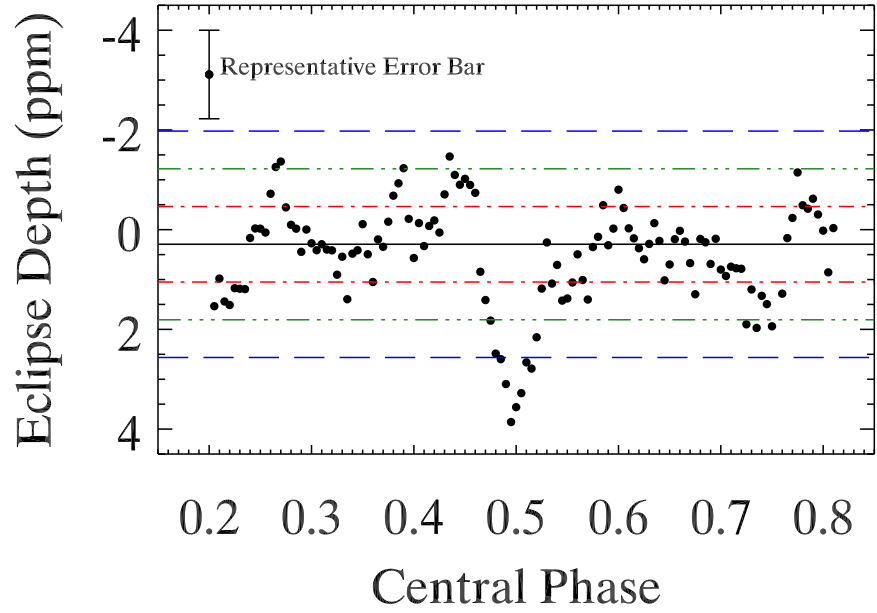}
\caption{Plotted are the measured eclipse depths vs. the phase where the eclipse model was centered.  Positive depths indicate an eclipse, while negative values indicate an inverted eclipse.  The y-axis is plotted in reverse to remind the eye that the detection is an eclipse.  The black solid line marks the mean depth, outside of the range of 0.45 to 0.55 in phase.  The red dash-dot lines mark $\pm$ one standard deviation about the mean, the green dash-dot-dot-dot line marks 2 standard deviations, and the blue dashed line marks 3 standard deviations.  The eclipse depth we measure at phase 0.5 stands out clearly.}
\label{fig:shift}
\end{figure}

We find generally low albedos across the groups.  The determined albedos for each radius group are not statistically secure measurements, since the 3$\sigma$ uncertainty range includes zero.  We do not force the albedos to be positive, nor do we force the eclipse depths to be positive.  The fact that we measure eclipses, rather than inverted eclipses, across the groups suggests that we are indeed seeing reflected light from the planets, though that reflected signal is weak.  To test this, we have looked for eclipses away from phase 0.5 using 49 of the candidates in our shortened 1 to 6 \rearth~group with the longest periods.  We chose the longest periods to provide ample room for normalizing the out-of-eclipse region while avoiding the transit.  To reproduce our method as faithfully as possible, we introduced a phase offset to the data before applying our normalization and averaging codes.  This means that the codes still operate as though the data are centered at phase 0.5, but the actual phase being searched is different.  We measured depths for phases 0.205 to 0.810 in steps of 0.005.  Figure \ref{fig:shift} shows the measured eclipse depths versus the actual phase that the data (and therefore the eclipse model) are centered on.  We then take the mean and standard deviation of the depths, outside of the phase range 0.45 to 0.55, for comparison to the depth we find at phase 0.5.   There is a clear signal at phase 0.5, so we are in fact measuring reflected light.  We originally performed this test with 50 candidates, including Kepler-4b, but the significant eclipse detected at phase 0.7 for Kepler-4b, noted in section 3.2.1, affected the results.  Since the signal for Kepler-4b is plausibly an eclipse, we removed Kepler-4b from the test.

The super-Earths in our sample are sufficiently hot that they likely have, at most, a tenuous atmosphere dominated by vaporized sodium \citep[e.g.][]{schaefer,miguel,castan,kite}.  We may in some cases be seeing bare surfaces.  While we have determined that super-Earths, on average, appear to have low albedos, there is a subset of higher-albedo planets \citep{demory}, including Kepler-10b \citep{kep10}, which appear to be bright.  The low-albedo hot super-Earths likely have no clouds present, and the measured albedo is of the surface.  Three potential scenarios could explain the subset of higher albedos: embedded higher-albedo particulates in a lava ocean surface, clouds, and thermal inversions.  The lava ocean scenario proposed by \citet{leger2011} for CoRoT-7b would have a Bond albedo less than 0.1 if made mostly of pure Al$_2$O$_3$, but \citet{rouan} propose that solid particles of ThO$_2$ floating in the lava ocean, depending on particle size, could produce an overall Bond albedo for the lava ocean of nearly 0.50, which could help to explain the unusual brightness of Kepler-10b.  Also, \citet{kite} note that as the CaO/SiO$_2$ ratio increases, there seems to be a rise in reflectivity in the UV and visible.  The data on this effect, however, are still very limited, but this could explain the higher albedo surfaces.  Alternatively, under certain conditions, the tenuous sodium atmospheres could support the formation of silicate clouds \citep[e.g.][]{schaefer,miguel,castan}, which would have higher albedos.  Cloud properties and coverage would depend on many factors, including composition, particle size, and winds, which would explain the existence of the bright and dark populations in similar temperature regimes.  Silicates in the atmosphere could also absorb UV and visible light and lead to a thermal inversion \citep{ito}, producing emission lines that would then make the secondary eclipse deeper than expected from just the combination of reflected light and thermal emission at the equilibrium temperature.  

The mini-Neptunes and super-Neptunes should have substantial atmospheres.  Hot Jupiters were expected to show a range of albedos depending on temperature \citep{sudarsky}, since temperature would control the type of clouds that could form and their location in the atmosphere.  A range has indeed been observed.  \citet{tres2b} used the amplitude of the phase curve in {\it Kepler} data for TrES-2b to determine that the geometric albedo, if only due to reflected light, is 0.0253 $\pm$ 0.0072.  The authors conclude that the true geometric albedo must be even lower, less than 1\%, since there is a significant thermal component.  Other low-albedo hot Jupiters from the mission include Kepler-423b ($A_g$ = 0.055 $\pm$ 0.028, \citet{k423b}), Kepler-12b ($A_g$ = 0.14 $\pm$ 0.04, \citet{k12b}), and HAT-P-7b ($A_g \lesssim$ 0.03, \citet{morris}), and others listed in Table \ref{tab:ecltab}.  Some hot Jupiters have been found to be bright, however, even at the same equilibrium temperatures as the dark planets.  Kepler-7b and Kepler-12b are good examples of the complex nature of cloud formation and properties.  Despite both planets receiving similar amounts of light from their stars, Kepler-7b is relatively bright, with $A_g \approx 0.35$ and shows evidence for cloud coverage that varies with longitude \citep{demoryetal};  Meanwhile Kepler-12b is dark, with $A_g = 0.08-0.14$ and shows no evidence for longitudinal variation \citep{k12b,hengdemory}.  Table \ref{tab:ecltab} includes other bright hot Jupiters, such as Kepler-13Ab ($A_g = 0.33^{+0.04}_{-0.06}$ \citet{shporer}).  Further examples of detected secondary eclipses of hot Jupiters in {\it Kepler} data can be found in \citet{angerhausen} and \citet{coughlin}.  A sample of these albedos have been included in Figure \ref{fig:alb} for reference, along with selected Solar System values\footnote{Radii for Solar System planets from NASA's Planetary Fact Sheet - Ratio to Earth Values at \burl{http://nssdc.gsfc.nasa.gov/planetary/factsheet/planet_table_ratio.html}, curated by Dr. David R. Williams.  Solar system albedos are from \citet{ssalb}.}.

The consistency of the average albedo across the three size groups considered here, as well as with the hot Jupiters studied in the literature, suggests that most of these hot planets have minimal or no cloud coverage.  The known outliers in albedo, however, demonstrate the complex and important role of cloud formation.

\section{Summary}
We apply the averaging technique used in \citet{paperi} to long cadence {\it Kepler} data, for three radius groups:  1-2 \rearth, 2-4 \rearth, and 4-6 \rearth.  We find that all three groups are similarly dark, with average geometric albedos of 0.11 $\pm$ 0.06, 0.05 $\pm$ 0.04, and 0.11 $\pm$ 0.08, respectively, if heat is completely redistributed.  In the case of instantaneous re-radiation, there is a larger thermal contribution to the eclipse depths, making the geometric albedo even lower.  The albedo results are summarized in Table \ref{tab:albresults}.  These average albedos are slightly lower than the average albedo we found using short cadence data, and are similar to some of the dark hot Jupiters.  As with bright hot Jupiters like Kepler-7b, there are outliers at higher albedo for super-Earths, too.  The simplest solution therefore is that a similar mechanism may be at work to create the outliers across the range in radius.  Such a mechanism may be silicate clouds, which are postulated for high temperature atmospheres across a wide range of pressures \citep{silicates}.  Given the low average albedo, though, these clouds must be somewhat uncommon.

\section{Acknowledgements}
We thank the anonymous referee for their suggestions and comments which greatly improved this manuscript overall, and also in particular the derivation and discussion of the effects of dilution in section 5.1. This research was supported by NASA Astrophysics Data Analysis Program grant NNX15AE53G.  This paper includes data collected by the Kepler mission. Funding for the Kepler mission is provided by the NASA Science Mission directorate.  All of the data presented in this paper were obtained from the Mikulski Archive for Space Telescopes (MAST). STScI is operated by the Association of Universities for Research in Astronomy, Inc., under NASA contract NAS5-26555. Support for MAST for non-HST data is provided by the NASA Office of Space Science via grant NNX13AC07G and by other grants and contracts.  This research has made use of the NASA Exoplanet Archive, which is operated by the California Institute of Technology, under contract with the National Aeronautics and Space Administration under the Exoplanet Exploration Program.  

\facility{Kepler}

\clearpage

\clearpage

\clearpage

\clearpage

\clearpage

\clearpage

\clearpage

\begin{deluxetable}{lccccccc}
\tabletypesize{\footnotesize}
\tablecolumns{8}
\tablewidth{0pt}
\tablecaption{Candidate Parameters \label{tab:param1}}
\tablehead{
\colhead{KOI} & \colhead{\# eclipses}\tablenotemark{a} & \colhead{$R_p$\tablenotemark{b}} & \colhead{$\left(R_p/a\right)^2$\tablenotemark{b}} & \colhead{$a/R_*$\tablenotemark{b}} & \colhead{Max $T_{eq}$\tablenotemark{c}} & \colhead{Min $T_{eq}$\tablenotemark{d}} & \colhead{$T_{eff}$} \\
& & \colhead{$\left(R_{\earth}\right)$} & \colhead{(ppm)} & & \colhead{(K)} & \colhead{(K)} & \colhead{(K)}
}
\startdata
\tableline
\multicolumn{8}{l}{Kepler-10b} \\
\tableline
K00072.01 & 814 (792) &  1.45 &  13.51 &    3.4 & 2748.6 & 1209.6 &   5627 \\
\tableline
\multicolumn{8}{l}{super-Earths} \\
\tableline
K00191.03\tablenotemark{e} & 1315 (0) &  1.25 &  12.60 &    3.7 & 2674.6 & 1177.0 &   5700 \\
K00739.01 & 1005 (957) &  1.42 &  10.46 &    7.9 & 1201.1 &  528.6 &   3733 \\
K00936.02 & 1296 (1193) &  1.22 &  13.77 &    6.8 & 1236.9 &  544.3 &   3581 \\
K00952.05\tablenotemark{e} & 1024 (0) &  1.88 &  38.52 &    5.5 & 1429.6 &  629.1 &   3727 \\
K01050.01 & 866 (788) &  1.78 &  13.41 &    5.8 & 1915.8 &  843.1 &   5095 \\
K01169.01\tablenotemark{e} & 1829 (0) &  1.29 &  12.56 &    3.6 & 2691.3 & 1184.3 &   5676 \\
K01202.01 & 1385 (1311) &  1.47 &  13.56 &    5.1 & 1962.3 &  863.5 &   4894 \\
K01239.01 & 1210 (1160) &  1.74 &  18.35 &    3.6 & 2903.7 & 1277.8 &   6108 \\
K01300.01 & 1828 (1673) &  1.10 &  16.31 &    4.7 & 1849.5 &  813.9 &   4441 \\
K01367.01 & 2159 (1985) &  1.40 &  23.12 &    3.7 & 2394.6 & 1053.8 &   5076 \\
K01424.01 & 1071 (1026) &  1.58 &  10.88 &    6.1 & 1766.0 &  777.1 &   4845 \\
K01442.01 & 1379 (1291) &  1.21 &  11.49 &    3.1 & 2904.2 & 1278.0 &   5626 \\
K01475.01 & 717 (671) &  1.81 &  11.95 &    8.6 & 1252.3 &  551.1 &   4056 \\
K01510.01 & 1485 (1412) &  1.70 &  20.22 &    4.7 & 2052.6 &  903.3 &   4924 \\
K01655.01 & 1108 (1071) &  1.48 &  11.48 &    4.0 & 2651.6 & 1166.8 &   5902 \\
K01662.01\tablenotemark{e} & 1319 (0) &  1.66 &  18.81 &    4.2 & 2401.8 & 1056.9 &   5463 \\
K01875.02 & 1691 (1595) &  1.31 &  19.00 &    3.0 & 3119.6 & 1372.8 &   5953 \\
K01880.01 & 1041 (940)&  1.56 &  14.41 &    7.1 & 1307.1 &  575.2 &   3855 \\
K01981.01 & 900 (876) &  1.61 &  11.99 &    4.3 & 2307.6 & 1015.5 &   5265 \\
K02119.01 & 2032 (1904) &  1.38 &  20.76 &    3.2 & 2625.0 & 1155.1 &   5203 \\
K02223.01 & 1104 (1070) &  1.64 &  13.80 &    5.4 & 1950.4 &  858.3 &   5002 \\
K02250.02 & 1040 (985) &  1.71 &  29.98 &    3.5 & 2385.5 & 1049.8 &   4922 \\
K02266.01 & 827 (782) &  1.70 &  17.12 &    5.1 & 1984.1 &  873.1 &   4949 \\
K02325.01\tablenotemark{e} & 1181 (0) &  1.35 &  10.09 &    4.9 & 2050.7 &  902.4 &   5035 \\
K02350.01 & 1132 (1109)&  1.75 &  14.46 &    4.3 & 2325.8 & 1023.5 &   5325 \\
K02355.01 & 1047 (1007)&  1.64 &  12.07 &    5.7 & 1837.2 &  808.5 &   4859 \\
K02393.02\tablenotemark{e} & 1247 (0) &  1.11 &  10.34 &    4.5 & 2082.2 &  916.3 &   4894 \\
K02396.01 & 2287 (2037) &  1.88 &  44.52 &    2.8 & 3008.5 & 1323.9 &   5529 \\
K02409.01 & 1732 (1604) &  1.47 &  25.49 &    3.7 & 2469.3 & 1086.6 &   5256 \\
K02453.01\tablenotemark{e} & 778 (0) &  1.54 &  11.31 &   10.5 &  994.8 &  437.8 &   3565 \\
K02480.01\tablenotemark{e} & 1875 (0) &  1.31 &  20.57 &    4.8 & 1649.8 &  726.0 &   3990 \\
K02493.01 & 1537 (1446) &  1.48 &  19.70 &    3.9 & 2375.5 & 1045.4 &   5166 \\
K02589.01 & 1920 (1816) &  1.19 &  12.56 &    3.8 & 2388.7 & 1051.2 &   5177 \\
K02607.01\tablenotemark{e} & 1620 (0) &  1.49 &  15.53 &    3.9 & 2697.9 & 1187.2 &   5883 \\
K02668.01 & 1321 (1227) &  1.25 &  13.67 &    4.0 & 2534.2 & 1115.2 &   5596 \\
K02694.01 & 1422 (1360) &  1.57 &  16.83 &    4.1 & 2151.2 &  946.6 &   4818 \\
K02699.01 & 2069 (1916) &  1.51 &  24.85 &    3.5 & 2514.9 & 1106.7 &   5216 \\
K02708.01 & 1438 (1366) &  1.71 &  19.72 &    4.8 & 1967.6 &  865.9 &   4790 \\
K02716.01 & 1281 (1218) &  1.48 &  11.86 &    4.9 & 2322.0 & 1021.8 &   5693 \\
K02735.01\tablenotemark{e} & 2195 (0) &  1.39 &  21.39 &    3.5 & 2507.3 & 1103.3 &   5154 \\
K02763.01\tablenotemark{e} & 2337 (0) &  1.13 &  18.46 &    3.4 & 2354.9 & 1036.3 &   4787 \\
K02796.01\tablenotemark{f} & 2262 (0) &  1.01 &  10.62 &    2.8 & 3270.6 & 1439.2 &   6108 \\
K02797.01 & 1275 (1243) &  1.80 &  16.28 &    3.7 & 2904.4 & 1278.1 &   6173 \\
K02817.01 & 956 (895) &  1.45 &  20.62 &    3.9 & 2399.8 & 1056.0 &   5238 \\
K02852.01 & 1600 (1498) &  1.44 &  16.07 &    3.4 & 2913.1 & 1281.9 &   5966 \\
K02882.02 & 1528 (1381) &  1.04 &  17.46 &    3.7 & 2094.6 &  921.8 &   4467 \\
K02886.01 & 1250 (1184)&  1.28 &  10.16 &    4.7 & 2186.0 &  961.9 &   5260 \\
K03089.01\tablenotemark{e} & 1284 (0) &  1.36 &  12.03 &    4.2 & 2335.0 & 1027.5 &   5269 \\
K03246.01 & 1126 (1118) &  1.86 &  27.52 &    1.9 & 3218.6 & 1416.4 &   4854 \\
K03867.01 & 1390 (1346) &  1.55 &  12.20 &    4.4 & 2508.0 & 1103.7 &   5825 \\
K04002.01\tablenotemark{e} & 1609 (0) &  1.42 &  23.04 &    2.9 & 2856.2 & 1256.9 &   5396 \\
K04109.01 & 1450 (1436) &  1.53 &  20.19 &    1.8 & 3367.4 & 1481.9 &   4968 \\
K04325.01 & 2017 (1908) &  1.19 &  12.39 &    3.2 & 2993.9 & 1317.5 &   5936 \\
K04512.01\tablenotemark{e} & 1800 (0) &  1.07 &  13.50 &    3.4 & 2576.4 & 1133.8 &   5286 \\
K04595.01\tablenotemark{e} & 2081 (0) &  1.25 &  18.73 &    4.2 & 2203.3 &  969.6 &   4985 \\
\enddata
\tablenotetext{a}{Listed are the number of eclipses used for the group based on the 2015 February catalog.  In parenthesis are the number of eclipses used for the shortened group.}
\tablenotetext{b}{$R_p$, $a$, and $R_*$ are taken from the Exoplanet Archive candidate table, and $(R_p/a)^2$ and $a/R_*$ are calculated from them.}
\tablenotetext{c}{Assumes $f = 2/3$ (instant re-radiation) and $A_B = 0.0$.}
\tablenotetext{d}{Assumes $f=1/4$ (complete redistribution) and $A_B = 0.9$ (i.e. $A_g = 0.6$).}
\tablenotetext{e}{Has false positive probability \citep{fpp} $>$ 1\%.}
\tablenotetext{f}{Is now listed as a false positive in the cumulative candidate table due to a non-transit-like shape.}
\end{deluxetable}

\begin{deluxetable}{lccccccc}
\tabletypesize{\footnotesize}
\tablecolumns{8}
\tablewidth{0pt}
\tablecaption{Candidate Parameters \label{tab:param2}}
\tablehead{
\colhead{KOI} & \colhead{\# eclipses}\tablenotemark{a} & \colhead{$R_p$\tablenotemark{b}} & \colhead{$\left(R_p/a\right)^2$\tablenotemark{b}} & \colhead{$a/R_*$\tablenotemark{b}} & \colhead{Max $T_{eq}$\tablenotemark{c}} & \colhead{Min $T_{eq}$\tablenotemark{d}} & \colhead{$T_{eff}$} \\
& & \colhead{$\left(R_{\earth}\right)$} & \colhead{(ppm)} & & \colhead{(K)} & \colhead{(K)} & \colhead{(K)}
}
\startdata
\tableline
\multicolumn{8}{l}{mini-Neptunes} \\
\tableline
K00102.01 & 583 (576) &  3.76 &  31.79 &    5.0 & 2324.7 & 1023.0 &   5731 \\
K00104.01 & 461 (438) &  2.49 &  11.63 &   11.0 & 1156.4 &  508.9 &   4238 \\
K00191.02\tablenotemark{e} & 383 (0) &  2.79 &  12.14 &    8.4 & 1773.9 &  780.6 &   5700 \\
K00220.01 & 525 (521) &  3.60 &  19.98 &    8.9 & 1695.6 &  746.2 &   5588 \\
K00439.01 & 676 (656) &  3.92 &  32.24 &    7.7 & 1774.8 &  781.0 &   5438 \\
K00496.01 & 796 (764) &  2.54 &  17.58 &    5.2 & 2145.3 &  944.1 &   5417 \\
K00517.01 & 429 (420) &  3.64 &  16.30 &    7.1 & 1955.2 &  860.4 &   5749 \\
K00526.01 & 619 (608) &  2.96 &  14.50 &    7.3 & 1908.7 &  839.9 &   5705 \\
K00676.02\tablenotemark{e} & 436 (0) &  2.23 &  10.58 &   11.9 & 1027.4 &  452.1 &   3914 \\
K00697.01 & 205 (205) &  3.49 &  12.83 &    5.7 & 2292.6 & 1008.9 &   6042 \\
K00732.01 & 944 (895) &  3.69 &  45.88 &    5.2 & 2193.5 &  965.2 &   5546 \\
K00780.01 & 551 (529) &  2.38 &  10.22 &    9.4 & 1472.0 &  647.8 &   4989 \\
K00800.01 & 472 (462) &  3.62 &  14.93 &    8.1 & 1962.1 &  863.4 &   6167 \\
K00844.01 & 271 (267) &  3.96 &  13.33 &   11.7 & 1475.3 &  649.2 &   5576 \\
K00916.01 & 395 (387) &  3.63 &  12.75 &    9.9 & 1613.6 &  710.1 &   5609 \\
K00926.01 & 401 (401) &  3.97 &  15.32 &    9.3 & 1778.9 &  782.8 &   6007 \\
K00941.02 & 382 (376) &  2.67 &  11.66 &    7.4 & 1727.8 &  760.3 &   5188 \\
K01357.01\tablenotemark{e} & 356 (0) &  3.09 &  10.93 &    9.0 & 1570.4 &  691.1 &   5209 \\
K01393.01\tablenotemark{e} & 650 (0) &  2.55 &  22.10 &    8.8 & 1177.7 &  518.3 &   3872 \\
K01428.01 & 1360 (1252) &  2.10 &  27.68 &    4.7 & 2053.5 &  903.6 &   4911 \\
K01557.01 & 268 (257) &  3.80 &  15.81 &    9.8 & 1416.2 &  623.2 &   4910 \\
K01762.01\tablenotemark{e} & 1480 (0) &  2.24 &  29.72 &    4.3 & 2619.3 & 1152.6 &   5985 \\
K01835.02 & 338 (337) &  2.55 &  10.90 &    4.4 & 2225.9 &  979.5 &   5192 \\
K01845.01 & 615 (609) &  3.35 &  22.17 &    3.9 & 2231.5 &  982.0 &   4883 \\
K01988.01 & 1312 (1280)&  3.48 &  64.88 &    2.2 & 2907.0 & 1279.2 &   4777 \\
K02034.01 & 228 (226) &  3.47 &  10.55 &   11.3 & 1525.3 &  671.2 &   5668 \\
K02104.01 & 590 (551) &  3.11 &  15.45 &    6.5 & 2174.7 &  957.0 &   6153 \\
K02269.01\tablenotemark{e} & 1850 (0) &  2.78 &  57.60 &    2.6 & 3634.2 & 1599.3 &   6517 \\
K02715.02 & 271 (262) &  3.26 &  21.13 &    9.2 & 1306.3 &  574.8 &   4385 \\
K02795.01\tablenotemark{e} & 424 (0) &  3.87 &  18.04 &    7.2 & 2105.6 &  926.6 &   6237 \\
K02842.01 & 310 (292) &  2.61 &  29.11 &    9.6 & 1015.7 &  447.0 &   3485 \\
K03913.01\tablenotemark{e} & 2126 (0) &  2.53 &  59.23 &    3.0 & 3275.9 & 1441.6 &   6263 \\
K03984.01 & 858 (849) &  2.20 &  14.62 &    3.6 & 2513.6 & 1106.1 &   5305 \\
K04098.01\tablenotemark{e} & 814 (0) &  2.38 &  20.85 &    1.9 & 3330.4 & 1465.6 &   5023 \\
K04561.01 & 58 (50) &  2.34 &  14.25 &    4.5 & 2692.0 & 1184.6 &   6302 \\
K04928.01 & 85 (80) &  2.27 &  16.50 &   27.1 &  557.7 &  245.4 &   3212 \\
K05566.01 & 30 (30) &  3.19 &  37.45 &    1.7 & 4445.5 & 1956.3 &   6456 \\
K05717.01\tablenotemark{e} & 125 (0) &  3.94 &  60.91 &    6.1 & 1689.1 &  743.3 &   4611 \\
\tableline
\multicolumn{8}{l}{super-Neptunes} \\
\tableline
K00007.01\tablenotemark{f} & 258 (0) &  4.14 &  16.13 &    6.2 & 2104.8 &  926.2 &   5781 \\
K00046.01 & 350 (349) &  4.36 &  15.74 &    8.3 & 1803.3 &  793.6 &   5761 \\
K00141.01\tablenotemark{e} & 430 (0) &  5.14 &  34.63 &    9.1 & 1610.4 &  708.7 &   5377 \\
K00221.01 & 371 (367) &  4.55 &  22.78 &   12.2 & 1377.7 &  606.2 &   5332 \\
K00240.01 & 303 (296) &  4.07 &  10.62 &   11.1 & 1688.4 &  743.0 &   6225 \\
K00242.01 & 177 (175) &  5.70 &  10.88 &   17.2 & 1228.1 &  540.4 &   5638 \\
K00433.01 & 261 (259) &  4.86 &  17.99 &   11.7 & 1467.9 &  646.0 &   5551 \\
K00470.01 & 273 (268) &  4.74 &  17.40 &   11.0 & 1576.6 &  693.8 &   5776 \\
K00531.01\tablenotemark{e} & 322 (0) &  5.69 &  36.88 &   14.3 &  962.8 &  423.7 &   4030 \\
K00766.01 & 307 (306) &  4.46 &  12.89 &   10.3 & 1737.0 &  764.4 &   6174 \\
K00782.01 & 196 (196) &  5.38 &  11.69 &   14.6 & 1418.4 &  624.2 &   5992 \\
K00851.01 & 262 (260) &  5.69 &  20.52 &   13.1 & 1450.6 &  638.3 &   5815 \\
K00953.01 & 349 (348) &  4.87 &  20.15 &   10.8 & 1579.6 &  695.1 &   5739 \\
K01815.01 & 354 (353) &  5.46 &  31.55 &    5.1 & 1943.8 &  855.4 &   4854 \\
K01845.02\tablenotemark{e} & 219 (0) &  5.99 &  20.24 &    7.3 & 1631.3 &  717.9 &   4883 \\
K02688.01 & 140 (139) &  5.20 &  19.31 &   18.3 &  874.4 &  384.8 &   4141 \\
\enddata
\tablenotetext{a}{Listed are the number of eclipses used for the groups based on the 2015 February catalog.  In parenthesis are the number of eclipses used for the shortened groups.}
\tablenotetext{b}{$R_p$, $a$, and $R_*$ are taken from the Exoplanet Archive candidate table, and $(R_p/a)^2$ and $a/R_*$ are calculated from them.}
\tablenotetext{c}{Assumes $f = 2/3$ (instant re-radiation) and $A_B = 0.0$.}
\tablenotetext{d}{Assumes $f=1/4$ (complete redistribution) and $A_B = 0.9$ (i.e. $A_g = 0.6$).}
\tablenotetext{e}{Has false positive probability \citep{fpp} $>$ 1\%.}
\tablenotetext{f}{Has secondary eclipse measured at phase $\sim$ 0.7.}
\end{deluxetable}

\begin{deluxetable}{lccccc}
\tabletypesize{\footnotesize}
\tablecolumns{6}
\tablewidth{0pt}
\tablecaption{Albedo Results \label{tab:albresults}}
\tablehead{
\colhead{Candidate} & \colhead{\# of} & \colhead{Total \# of} & \colhead{Eclipse Depth} & \colhead{$A_g$} & \colhead{$A_g$}\\
\colhead{Group} & \colhead{Candidates} & \colhead{Eclipses} & \colhead{(ppm)} & \colhead{(f = 1/4)} & \colhead{(f = 2/3)}
}
\startdata
\tableline
\multicolumn{6}{l}{Super-Earths}\\
\tableline
2015 February w/o Kepler-10b & 55 & 79,678 & 2.44 $\pm$ 0.99 & 0.11 $\pm$ 0.06 &  $<$ 0.17\tablenotemark{a} \\
Shortened, w/o Kepler-10b & 39 & 50,805 & 2.63 $^{+1.13}_{-1.14}$ & 0.11 $\pm$ 0.06 &  $<$ 0.17\tablenotemark{a} \\
2015 February w/ Kepler-10b & 56 & 80,492 & 3.56 $\pm$ 0.65 & 0.19 $\pm$ 0.04 & 0.04 $\pm$ 0.07 \\
Shortened, w/ Kepler-10b & 40 & 51,597 & 3.56 $\pm$ 0.67 & 0.19 $\pm$ 0.04 & 0.02$^{+0.07}_{-0.08}$ \\
\tableline
\multicolumn{6}{l}{Mini-Neptunes} \\
\tableline
2015 February & 38 & 22,677 & 2.42 $\pm$ 0.76 & 0.07 $\pm$ 0.03 & $<$ 0.04\tablenotemark{a} \\
Shortened & 28 & 13,580 & 1.69 $\pm$ 0.85 & 0.05 $\pm$ 0.04 & $<$ 0.07\tablenotemark{a} \\
\tableline
\multicolumn{6}{l}{Super-Neptunes}\\
\tableline
2015 February & 16 & 4,572 & 2.16$^{+1.37}_{-1.38}$ & 0.12 $\pm$ 0.08 & 0.09 $\pm$ 0.08 \\
Shortened & 12 & 3,316 & 4.67$^{+2.18}_{-2.15}$ & 0.23 $\pm$ 0.11 & 0.21 $\pm$ 0.11 %\\ 
\enddata
\tablenotetext{a}{3$\sigma$ upper limit.}
\end{deluxetable}

\begin{deluxetable}{lcccccc} % KOI, Confirmed Name, Radius, rpasq, eclipse depth, discovery?
\tabletypesize{\footnotesize}  %\small (11pt), \footnotesize (10pt), or \scriptsize (8pt).
\tablecolumns{7}
\tablewidth{0pt}
\tablecaption{Sample of Individual Detections of Secondary Eclipses in {\it Kepler} Data \label{tab:ecltab}}
\tablehead{
\colhead{KOI} & \colhead{Confirmed} & \colhead{$R_p$\tablenotemark{a}} & \colhead{$\left(R_p/a\right)^2$\tablenotemark{a}} & \colhead{Depth\tablenotemark{a}} &  \colhead{A$_g$} & \colhead{Source}  \\ 
& \colhead{Name} & \colhead{$\left(R_{\earth}\right)$} & \colhead{(ppm)} & \colhead{(ppm)} & &
}
\startdata
1.01 & TrES-2b & 13.68 & 251.73 & 7.7$\pm$2.6 & 0.031$\pm$0.010 & \citet{esteves2015} \\ 
2.01 & HAT-P-7b & 15.57\tablenotemark{b}& 349.5 & 69.1$\pm$3.8 & $\lesssim$ 0.03 & \citet{morris} \\ 
10.01 & Kepler-8b & 15.31 & 191.3 & 26.2$\pm$5.6 & 0.137$\pm$0.029 & \citet{esteves} \\ 
13.01 & Kepler-13Ab & 15.43 & 362.20 & 90.81$\pm$0.27 & 0.33$\pm$0.06 & \citet{shporer} \\ 
17.01 & Kepler-6b & 14.83 & 194.38 & 22$\pm$7 & 0.11$\pm$0.04 & \citet{56b} \\ 
18.01 & Kepler-5b & 16.04 & 182.87 & 21$\pm$6 & 0.12$\pm$0.04 & \citet{56b} \\ 
20.01 & Kepler-12b & 18.60 & 202.96 & 31$\pm$8 & 0.14$\pm$0.04 & \citet{k12b} \\ 
72.01 & Kepler-10b & 1.42 & 12.86 & 5.8$\pm$2.5 & 0.61$\pm$0.17 & \citet{kep10} \\ 
69.01 & Kepler-93b & 1.51 & 2.45 & 2.2$\pm$0.8 & 0.87-0.88 & \citet{demory} \\ 
97.01 & Kepler-7b & 17.71 & 145.83 & 44$\pm$5 & 0.32$\pm$0.03 & \citet{demoryetal} \\ 
135.01 & Kepler-43b & 13.17 & 153.01\tablenotemark{c}& 17.0$\pm$5.3 & 0.06$\pm$0.02 & \citet{angerhausen} \\
183.01 & Kepler-423b & 13.08 & 241.58 & 14.2$\pm$6.6 & 0.055$\pm$0.028 & \citet{k423b} \\ 
196.01 & Kepler-41b & 9.77 & 193.74 & 64$\pm$12 & 0.30$\pm$0.07 & \citet{k41b} \\ 
202.01 & Kepler-412b & 14.54 & 457.39 & 47.4$\pm$7.4 & 0.013-0.094 & \citet{k412b} \\  
1169.01 & \ldots & 1.26 & 19.2 &13.5$\pm$3.1 & 0.48-0.67 & \citet{demory} \\ 
1658.01 & Kepler-76b & 13.72 & 535.49\tablenotemark{c}& 75.6$\pm$5.6 & 0.22$\pm$0.02 & \citet{angerhausen} \\ 
2133.01 & Kepler-91b & 15.00 & 76.35 & 35$\pm$18 & 0.46$\pm$0.30 & \citet{esteves2015} \\ 
\ldots\tablenotemark{d} & Kepler-78b & 1.16 & 22.4 & 10.5$\pm$1.2 & 0.27-0.4\tablenotemark{e} & \citet{sanchis} 
\enddata 
\tablenotetext{a}{Values from the source paper unless otherwise noted.  Asymmetric error bars are given as symmetric with the larger of the two uncertainties.}
\tablenotetext{b}{Radius value from \citet{esteves2015}.}
\tablenotetext{c}{\rpasq~calculated from \citet{esteves2015}.}
\tablenotetext{d}{Kepler-78b does not have a KOI number, because it was not initially detected by the {\it Kepler} mission pipeline.  It was discovered by an independent search of the data reported in \citet{sanchis}.}
\tablenotetext{e}{$A_g$ estimated using $A_g = (2/3)A_B$ from $A_B$ given by the source paper.}
\end{deluxetable}

\end{document}